\documentclass[useAMS]{mn2e}
\usepackage{epsfig,natbib_vw,times}

\bibpunct[, ]{(}{)}{;}{a}{}{,}

\def \aj {AJ}
\def \mnras {MNRAS}
\def \apj {ApJ}
\def \apjs {ApJS}
\def \apjl {ApJL}
\def \aap {A\&A}

\def \araa {ARAA}

\def \be {\begin{equation}}
\def \ee {\end{equation}}

\def\gsim{\mathrel{\lower0.6ex\hbox{$\buildrel {\textstyle >}
 \over {\scriptstyle \sim}$}}}
\def\lsim{\mathrel{\lower0.6ex\hbox{$\buildrel {\textstyle <}
 \over {\scriptstyle \sim}$}}}

\def\m@th{\mathsurround=0pt }
\def\eqalign#1{\null\,\vcenter{\openup1\jot \m@th
 \ialign{\strut\hfil$\displaystyle{##}$&$\displaystyle{{}##}$\hfil
 \crcr#1\crcr}}\,}

\def \zabs {z_{\rm abs}}
\def \zem {z_{\rm em}}
\def \caii {Ca~{\sc ii}~}

\def \znii {Zn~{\sc ii}~}
\def \crii {Cr~{\sc ii}~}
\def \mnii {Mn~{\sc ii}~}
\def \tiii {Ti~{\sc ii}~}
\def \mgii {Mg~{\sc ii}~}
\def \feii {Fe~{\sc ii}~}
\def \mgi {Mg~{\sc i}~}
\def \civ {C~{\sc iv}~}
\def \hi {H~{\sc i}~}
\def \nhi {$N$(H~{\sc i})~}
\def \lognhi{$\log$[$N$(H~{\sc i})]~}
\def \EBV {\textit{E(B$-$V)}}

\title[Ca~{\sc ii}--selected DLAs]{Selecting damped Lyman-$\alpha$
  systems through \caii absorption. I: Dust depletions and reddening at
  $z\sim1$}
\author[V. Wild, P. Hewett \& M. Pettini]{Vivienne Wild$^{1,2}$\thanks{vwild@mpa-garching.mpg.de},
  Paul C. Hewett$^1$ and Max Pettini$^1$
\vspace*{6pt}\\
1. Institute of Astronomy, University of Cambridge, Cambridge CB3 0HA, UK \\
2. Max-Planck-Institut f\"{u}r Astrophysik, 85748 Garching, Germany\\}

\begin{document}
\maketitle
\begin{abstract}
  
  We use the average \EBV\ and \znii column densities of a sample of
  $z\sim1$ Ca~{\sc ii}~$\lambda\lambda 3935, 3970$ absorption line
  systems selected from the fourth data release of the Sloan Digital
  Sky Survey (SDSS)to show that on average, with conservative
  assumptions regarding metallicities and dust-to-gas ratios, they
  contain column densities of neutral hydrogen greater than the damped
  Lyman-$\alpha$ (DLA) limit.  We propose that selection by \caii
  absorption is an effective way of identifying high column densities
  of neutral hydrogen, and thus large samples of DLAs at
  $z_{abs}\lsim1.3$ from the SDSS.  The number density of strong \caii
  absorbers (with rest-frame equivalent width $W_{\lambda 3935} \geq
  0.5$\,\AA), is $\sim 20-30\%$ that of DLAs, after correcting for the
  significant bias against their detection due to obscuration of the
  background quasars by dust.  On average these absorbers have \EBV\,$
  \gsim 0.1$\,mag; the dustiest absorbers show depletions of
  refractory elements at a level of the largest depletions seen in
  DLAs.  For the first time we can measure the dust-to-metals ratio in
  a sample of absorption selected galaxies, and find values close to,
  or even larger than, those observed locally.  All of these
  properties suggest that a substantial fraction of the \caii
  absorbers are more chemically evolved than typical DLAs.  There is a
  trend of increasing dust content with $W_{\lambda3935}$; this trend
  with strong-line equivalent width is also observed in an equivalent,
  but much larger, sample of \mgii absorbers.  Such a trend would
  result if the dustier systems are hosted by more massive, or
  disturbed, galaxies.  Follow-up imaging is required to provide
  conclusive evidence for or against these scenarios.  From
  consideration of the \EBV\ distribution in our sample, and assuming
  \caii absorbers represent a subset of DLAs, we calculate that dust
  obscuration causes an underestimation in the number density of DLAs
  by at least $8 - 12$\% at these redshifts.  Finally, the removal of
  Broad Absorption Line (BAL) quasars from the SDSS quasar sample
  increases the sensitivity of the detection of reddening by
  intervening absorbers.  To this end, we describe a new, automated,
  principal component analysis (PCA) method for identifying BAL
  quasars.

\end{abstract}


\section{Introduction}

Measurements of the chemical composition of galaxies play an important
role in understanding galaxy formation and evolution.  The study of
galaxy metallicities is closely intertwined with the question of their
dust content: systems richer in metals have the greater potential to
form dust grains. These grains selectively deplete metals in the
interstellar medium (ISM) of the galaxies, redden their spectral
energy distribution and cause an overall extinction of the
light.  The uncertainty in dust composition and distribution in
galaxies affects the interpretation of many extragalactic
observations. 

Quasar absorption line systems provide a powerful probe of metals in
the ISM of galaxies at high redshift, unhampered by luminosity bias,
through the absorption of the light from a background quasar by gas
phase metal ions.  Notably, reddening due to dust has failed to be
detected in samples of the highest column density absorption line
systems, damped Lyman-$\alpha$ systems (DLAs), at $z\sim2-4$
\citep{2004MNRAS.354L..31M,corals_ebv}. The relative paucity of dust
particles in DLAs is backed up by observations of absorbers in radio
selected quasar spectra, which indicate that extinction due to dust is
a small effect \citep[CORALS survey][]{2001A&A...379..393E}, and metal
abundances show DLAs to be generally metal- and therefore dust-poor
\citep[e.g.][]{2004cmpe.conf..257P}.

\subsection{\caii in the local universe}
This paper investigates the dust and metal properties of a new class
of absorption line system, selected via the
\caii$\lambda\lambda3935,3970$ doublet\footnote{Vacuum wavelengths are
used throughout this paper.}  from the Sloan Digital Sky Survey
(SDSS). These are the familiar K and H lines of the solar spectrum.

Ca is an $\alpha$-capture element thought to be produced mainly by
Type II supernovae.  The convenient positioning of the resonance lines
of its singly-ionised ion in the violet portion of the optical
spectrum has made it accessible to stellar and interstellar studies
for over a century, with major Galactic surveys by
\citet{1949ApJ...109..354A} and \citet{1972ApJ...173...43M}.  The
inference of Ca abundance from \caii absorption lines is however
compromised by two considerations.  First, with an ionisation
potential of 11.9\,eV, lower than that of H~{\sc i}, Ca$^+$ is a minor
ionisation stage of Ca in the neutral ISM---most of the Ca is doubly
ionised.  Second, Ca is one of most depleted elements in the
ISM---typically more than 99\% of all Ca is `hidden from view' having
been incorporated into dust grains \citep{1996ARA&A..34..279S}. The
degree of depletion is expected to depend significantly on both the
density of the gas (not just $N$(H~{\sc i})) and on the presence of
shocks, which may result in grain destruction, returning Ca to the gas
phase. However, as stressed by \citet{2000ApJ...544L.107W}, the
dispersion in the measured gas-phase abundance of \caii at fixed N(HI)
in the Milky Way is remarkably small.  Although the precise interpretation of the
detection of a significant column density of \caii is complex, it is
clear that a large column density of neutral hydrogen is implied,
given the high degree of depletion and low ionisation energy.  

\caii absorption in quasar spectra due to low redshift galaxies was
detected early on in the history of quasar absorption line studies;
for a review of early results see \citet{1988qsal.proc..147B}. The
most comprehensive study was carried out by
\citet{1991MNRAS.249..145B} who concluded that at projected
separations $\gsim 20\,$kpc from the centres of visible galaxies,
\caii absorption is patchy and, where it does occur, relatively weak,
with the rest frame equivalent width of the strongest member of the
\caii doublet $W_{\lambda 3935}\lsim 0.2\,$\AA.  Examples of strong
\caii absorption ($W_{\lambda 3935}\gsim0.5\,$\AA) were mostly
confined to sightlines with projected separations $\lsim
10\,$kpc. Evidence for mergers was seen in those cases with large projected separations and large column
densities of \caii 
\citep{1991MNRAS.251..649B}.

Since the early 1990s the focus of quasar absorption line research moved
to high redshift and \caii was largely ignored due to both its rarity
and wavelength. At $z>1$ the doublet moves into the near infrared
which, at that time, was difficult to access.  Recently, very strong \caii
absorption has been detected in a small number of highly reddened,
\EBV$\sim 1$, sightlines to quasars
\citep{2000A&A...359..457P,2002ApJ...575L..51H,2005ApJ...622L.101W}.
Such detections confirm the link between high gas column densities,
dust and \caii absorption in relatively extreme circumstances when the
line-of-sight is almost coincident with the centre of a galaxy.

\subsection{Metal absorbers and DLAs}

DLAs are quasar absorption line systems with the highest column densities of
neutral hydrogen, with a nominal column density limit of \nhi
$>2\times10^{20}$cm$^{-2}$.  Although they have been studied
extensively since the seminal paper by \citet{1986ApJS...61..249W}
twenty years ago, their precise nature, evolution and relation to the
population of galaxies as a whole continue to be the subject of much
discussion \citep{wolfe05}. Interest in DLAs results primarily from their
apparent domination of the neutral gas mass fraction of the Universe
at high redshift. Such cold, neutral gas is required for later star formation.

The defining diagnostic of a DLA, the Lyman-$\alpha$ absorption line,
does not enter the optical atmospheric window until redshift $z_{\rm
abs} \sim 1.8$. At $z\sim 0$ DLAs can be identified through
21\,cm emission \citep[e.g.][]{zwaan05}. The need for space-based
observations has made it difficult to obtain large samples of
intermediate redshift DLAs which are crucial both to clarify the
relation of DLAs to luminous galaxies by direct imaging of the DLA
hosts (which is much easier at redshifts $z \lsim 1$), and to follow
the evolution of their properties over the course of time.

\citet{2000ApJS..130....1R} proposed a useful prognostic for
intermediate redshift DLAs: the equivalent widths of the strong
\mgii$\lambda2796$ and \feii$\lambda2600$ absorption lines can be used
to identify DLAs with a $\sim$40\% success rate
\citep{astro-ph/0505479}.  A complementary method to that proposed by
Rao \& Turnshek for recognising DLAs at intermediate redshift is
through the detection of absorption lines which are intrinsically
weak, due to low cosmic abundance, low transition probability,
ionisation state or affinity for dust grains.  By choosing lines
appropriately, the success rate of identification can approach 100\%,
circumventing to some extent the need for UV spectroscopy;
for example, the detection of the \znii$\lambda\lambda2026, 2062$
doublet virtually guarantees the DLA nature of an absorption system
\citep{1990ApJ...348...48P}.

As highlighted above, little is known of the strength of \caii lines
in DLAs; here we propose that approximately one fifth to one quarter
of DLAs have \caii lines with rest frame equivalent widths exceeding
0.5\,\AA\ for the stronger member of the doublet.  The Sloan Digital
Sky Survey (SDSS) quasar catalogue, with more than 45\,000 spectra in
its third data release \citep[DR3,][]{astro-ph/0503679}, provides the
ideal database for obtaining large samples of metal line systems over
a wide redshift range.  The ability to detect significant numbers of
DLAs using the SDSS over the entire redshift range $0\lsim\zabs<1.3$
would provide a much needed injection of new intermediate to low
redshift DLAs for imaging and chemical evolution studies.

\subsection{Aims of this paper}

In a recent paper \citep[][hereafter Paper I]{2005MNRAS.361L..30W}, we
assembled a sample of 31 \caii absorption systems selected from a high
signal--to--noise ratio subset of the SDSS DR3 at redshifts $0.84 < \zabs <1.3
$ and reported the detection of a significant degree of
reddening---corresponding to a colour excess \EBV$\, = 0.06$---of the
background quasars by dust in these absorbers.  In this follow-up
paper, we analyse the relative abundances of refractory elements in
these intermediate redshift \caii absorbers with a view to
establishing their connection to the DLA population.

Specifically, we construct composite spectra which allow us to measure
the relative abundances of Zn~{\sc ii}, Cr~{\sc ii}, Fe~{\sc ii},
Mn~{\sc ii}, Ti~{\sc ii} and Ca~{\sc ii}.  The Zn measurement is
crucial since, among the elements considered, it is the only one which
shows little affinity for dust \citep{1996ARA&A..34..279S} and
therefore gives a standard against which to compare the gas-phase
abundances of the others.  We add to the sample of Paper~I six
additional \caii systems identified from the fourth data release (DR4)
of the SDSS (Adelman-McCarthy et~al.  2005)\nocite{DR4}, taking the
total number of \caii absorbers to 37.  27 of these have rest frame
spectra that encompass the region of the \znii doublet. We also
re-examine in more detail the reddening results of Paper~I in light of
the abundance determinations presented here.  


The paper is organised as follows.  In Section 2 we describe our
search criteria for identifying absorption line systems.  Strong \mgii
$\lambda2796$ and \feii $\lambda2600$ absorption lines are seen in all
\caii absorbers; we compare their equivalent width distributions with
those of \mgii absorption systems at similar
redshifts in SDSS quasars, paying particular consideration to the DLA
selection criteria of \citet{astro-ph/0505479}.  In Section 3 we
calculate the redshift path of our \caii survey. In Section 4 we
derive column densities and relative abundances of elements detected
in composite spectra of \caii absorption line systems; we compare the
abundance patterns with those seen in known DLAs and in the
interstellar medium of the Milky Way.  Section 5 deals with the
reddening introduced by different samples of \caii and \mgii absorbers
on the observed spectral energy distributions (SEDs) of the background quasars.  We discuss our
results in Section 6, focusing in particular on the use of \caii in
the identification of DLAs, on the metallicities and dust content of
Ca~{\sc ii}-selected DLAs at $z \sim 1$, and on the implications of
our findings for the more general issue of quasar obscuration by
dust-rich intervening absorbers. Appendix A gives details of our new
method for recognising quasars that show evidence of broad absorption
line systems (BALs).

\section{\caii and \mgii absorption systems}\label{sec:sample}


\begin{table*}
  \begin{center}
    \caption{\label{tab:1} \small Name and spectroscopic
      identification of each quasar in our DR4 \caii absorber sample,
      together with rest frame equivalent widths of
      \caii$\lambda\lambda3935,3970$, \mgii$\lambda\lambda2796,2804$,
      \mgi$\lambda2853$ and \feii$\lambda2600$. The final column gives
      the estimated reddening of each quasar by the \caii absorber,
      calculated assuming an extinction curve similar to that of the
      Large Magellanic Cloud (see Section \ref{sec:red}).  See Table~1
      of Paper I for equivalent details of the DR3 \caii systems.}
    \begin{minipage}{18.5cm}
      \begin{tabular}{ccccccccccc} \hline\hline
        SDSS ID & MJD\footnote{Modified Julian date},plate,fibre &
        $i$\footnote{PSF $i$-band magnitude, corrected for Galactic
          extinction} & $\zem$ & $\zabs$\footnote{Measured from
          \mgii{$\lambda2796$}} & $W_{\lambda3935,3970}$ & err($W$) 
          & $W_{\lambda2796, 2803}$ & $W_{\lambda2853}$ 
          & $W_{\lambda2600}$ & \EBV \\ \hline
        J080958.56+515118.0 & 53090,1780,532 & 18.56 & 1.289 & 0.902 &  0.64, 0.39 &  0.13, 0.11 &  2.25, 2.22 &  1.14 &  1.92 & $-$0.007 \\
        J120300.96+063440.8 & 53089,1623,209 & 18.47 & 2.182 & 0.862 &  1.42, 0.89 &  0.29, 0.24 &  5.57, 4.97 &  2.98 &  3.75 &  0.417 \\
        J122756.40+425631.2 & 53112,1452,505 & 17.17 & 1.310 & 1.045 &  0.39, 0.17 &  0.07, 0.08 &  1.61, 1.39 &  0.25 &  1.15 &  0.026 \\
        J130841.28+133130.0 & 53089,1772,122 & 18.60 & 1.954 & 0.951 &  0.87, 0.91 &  0.27, 0.29 &  1.82, 1.61 &  0.68 &  1.16 &  0.080 \\
        J140444.16+551637.2 & 53088,1324,421 & 18.46 & 1.589 & 1.070 &  1.33, 0.54 &  0.34, 0.22 &  1.98, 2.11 &  0.94 &  1.44 &  0.195 \\
        J153503.36+311832.4 & 53119,1388,068 & 17.79 & 1.510 & 0.904 &  0.84, 0.36 &  0.15, 0.12 &  2.11, 1.80 &  0.82 &  1.03 &  $-$0.008 \\ 
        \hline
      \end{tabular}
\vspace*{-0.4cm}
    \end{minipage}
  \end{center}
\end{table*}


\begin{figure*}
  \hspace*{-0.5cm}
  \begin{center}
  \begin{minipage}{14cm}
    \includegraphics[scale=0.625]{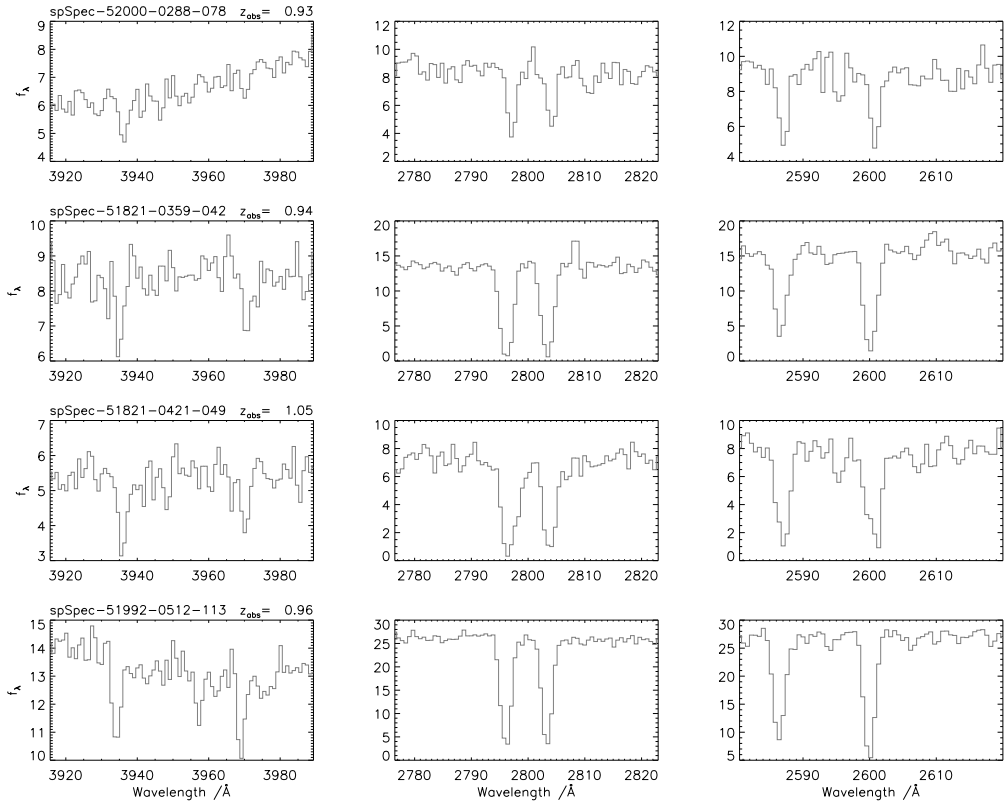}
  \end{minipage}
  \end{center}
  \caption{Examples of \caii systems (left panels) found in the SDSS quasar sample, along
    with the associated \mgii (centre) and \feii lines (right).  The
    units of the $y$-axis are $10^{-17}$\,erg~cm$^{-2}$~s$^{-1}$~\AA$^{-1}$. 
    The SDSS spectroscopic filename (MJD, plate number, fibre number) 
    and absorber redshift are given above each panel in the left-hand column.}
   \label{fig:caegs}
\end{figure*}


\begin{figure*}
  \hspace*{-0.5cm}
  \begin{minipage}{\textwidth}
    \includegraphics[scale=0.75]{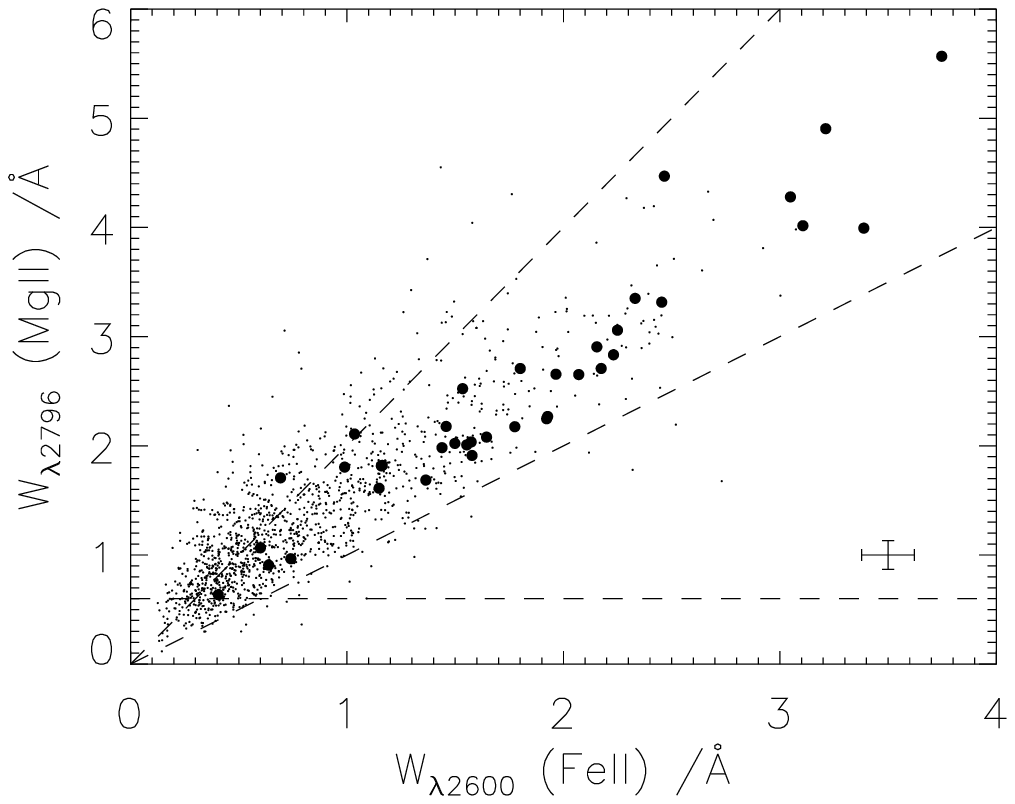}
    \includegraphics[scale=0.75]{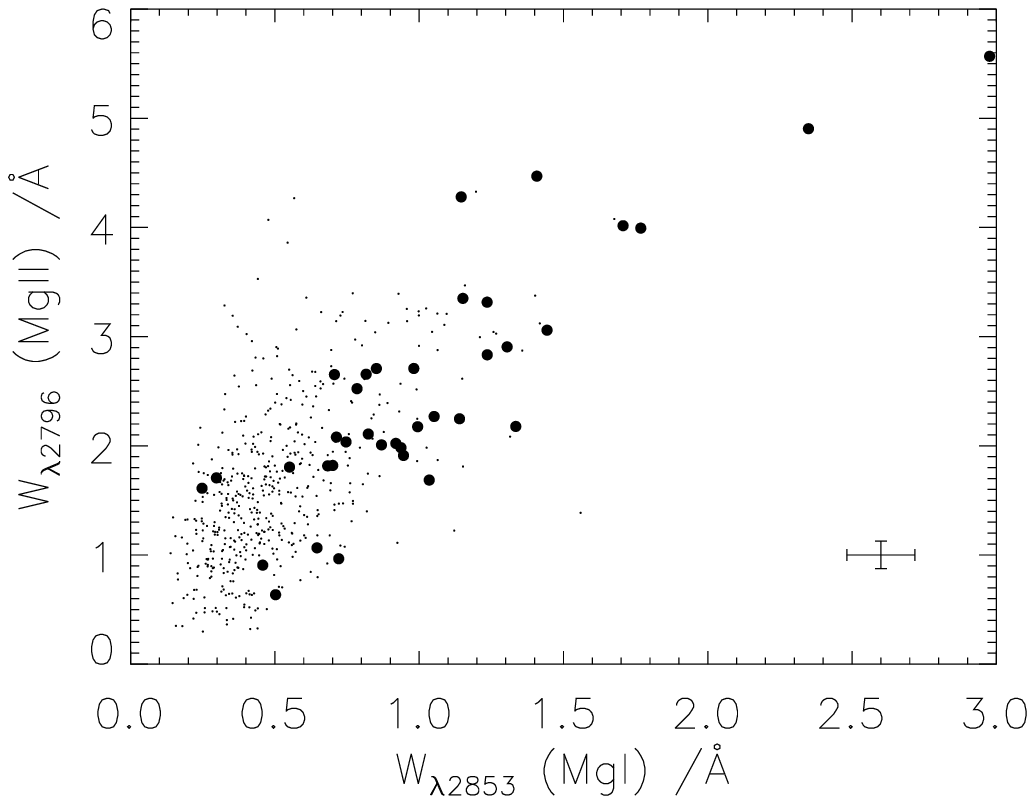}
  \end{minipage}
  \caption{The rest frame equivalent widths of \mgii$\lambda2796$ vs.
    \feii$\lambda2600$ on the left and \mgi$\lambda2853$ on the right
    for the \caii absorbers (filled symbols) and \mgii absorbers
    (small points).  The dashed lines in the left panel indicate the
    limits within which DLAs have been found in \mgii absorbers by
    \citet{astro-ph/0505479}.  The error bars in the bottom right-hand
    corners show the typical errors on the equivalent widths.}
  \label{fig:mgiifeii}
\end{figure*}

The sample of quasar spectra used during the absorption line search
and creation of control quasar composite spectra for the reddening
analysis was restricted to those with spectroscopic signal--to--noise
ratio (SNR)\,$> 10$ in the $i$-band and with Galactic
extinction-corrected point spread function (PSF) magnitudes $i < 19.0$
in order to minimise the number of false detections during the \caii
line searches.  The magnitude cut is very similar to that used in
constructing the main spectroscopic quasar sample from the SDSS
photometric catalogue ($i \leq 19.1$).  After exclusion of BAL
quasars, the final sample consists of 11\,371 quasars from the DR3
quasar catalogue \citep{astro-ph/0503679} and a further 3\,153 from
the quasars observed on the 266 additional spectroscopic plates in
DR4.

We searched the 14\,524 quasars independently for both
\caii$\lambda\lambda3935, 3970$ and \mgii$\lambda\lambda2796, 2804$
absorption doublets at redshifts $0.84< \zabs <1.3$.  Residual sky OH
emission features were subtracted from all spectra using the method of
\citet{skysub}, which greatly reduces residual sky noise in the
wavelength region around the \caii lines.  A ``continuum'' was defined
for each quasar via the application of a simple 41--pixel median
filter.  The ``difference'' spectrum, to be searched for absorption
features, was then obtained by subtracting the continuum from the
original quasar spectrum.  The absorption line search used a
matched-filter technique \citep[e.g.][]{1985MNRAS.213..971H} with two
template Gaussian doublets of the appropriate wavelength separation
and full width at half maximum FWHM = 160\,km~s$^{-1}$ (the resolution
of the SDSS spectra), 200\,km~s$^{-1}$ and 240\,km~s$^{-1}$.  Three
values of the doublet ratio were incorporated into the search: 2:1
(corresponding to unsaturated lines on the linear part of the curve of
growth), 1:1 (for saturated lines on the flat part of the curve of
growth) and an intermediate case, 4:3.  At each pixel the template
giving the largest SNR was determined and candidate absorbers were
selected by applying a threshold value for the SNR (SNR$>6\sigma$ for
\mgii and SNR$>5\sigma$ for \caii).

The redshift range $0.84< \zabs <1.3$ was determined in Paper I from
consideration of, respectively, the lowest redshift at which the broad
2175\,\AA\ bump in the Milky Way reddening curve is detectable, and
the highest redshift at which the \caii doublet can be found in SDSS
spectra.  For the metal abundance analysis presented in this paper, we require
coverage of the \znii$\lambda\lambda2026, 2062$ lines associated with
the \caii systems; this increases the lower redshift limit to $\zabs >
0.88$.  We also exclude systems where the 2175\,\AA \ bump (or \znii
doublet) falls below the Lyman~$\alpha$ emission line to avoid
confusion with the Lyman~$\alpha$ forest.

\subsection{\caii systems}

At the redshifts of interest the \caii lines fall in the red portion
of the optical spectrum, beyond 7250\,\AA, where the SDSS quasar
spectra become progressively noisier due to the increasing sky
background and decreasing instrumental sensitivity.  Thus, we required
candidate \caii systems to possess corresponding \mgii absorption
within $\pm 200$~km~s$^{-1}$.  This list of \caii candidates was
subjected to further scrutiny by a fully parameterised fit of the
absorption lines.  A continuum was fitted to the regions around the
\mgii and \caii lines and the corresponding portions of the quasar
spectra were normalised by dividing by the continuum level.  Gaussian
doublets were then fitted to the normalised spectra using a
maximum-likelihood routine; the position and line width of the
doublet, and the relative strengths of the individual lines, were
allowed to vary freely.  Rest frame equivalent widths ($W$) were
calculated from the parameters of the Gaussian fits and errors
estimated by propagation of the parameter errors derived during the
maximum likelihood fit.  For the purposes of composite construction
(Sections \ref{sec:comp1} and \ref{sec:qsorefs}) absorption redshifts
were redefined using the centre of the fitted \mgii$\lambda 2796$
line.  A small number of \caii doublets for which the fit proved to be
visually unsatisfactory were removed from the sample.

Our final catalogue consists of 37 \caii systems with $W_{\lambda
3935} \gsim 0.35$\,\AA\ and $\langle \zabs \rangle = 0.95$; 27 of
these are suitable for use in the \znii analysis, having $\zabs>0.88$.
Fig.~\ref{fig:caegs} shows some examples of spectra with \caii systems
in regions around the \caii$\lambda\lambda3935, 3970$,
\mgii$\lambda\lambda2796, 2804$ and \feii$\lambda\lambda2587, 2600$
doublets.  The rest frame equivalent widths of the \caii and \mgii
doublets, together with \feii$\lambda2600$ and \mgi$\lambda2853$ for
all the \caii systems in the DR3 quasar catalogue can be found in
Table~1 of Paper I.  Table~\ref{tab:1} of this paper gives the same
information for the additional six \caii systems found in DR4 quasar
spectra.

\subsection{\mgii systems}\label{sec:mgii}

It is interesting to compare the properties of the rare \caii
absorbers with those of other classes of quasar absorption line
systems.  To this end, a catalogue of 2\,338 \mgii absorbers was
compiled from the sample used for confirmation of the DR3 \caii
absorption systems. The same fully parameterised fit was carried out
as described above for the \caii sample. However, because of the large
number of systems, we automatically culled those absorbers from the
sample with poor continuum fits, unphysical line ratios and detection
significances (for the \mgii doublet) of less than 4, rather than
relying on visual inspection of individual systems. \mgii absorbers
also appearing in our \caii catalogue were removed from this sample.

Among these \mgii systems we are especially interested in the subset
satisfying the criteria for likely DLA candidates.  Specifically,
\citet{astro-ph/0505479} found that 43\% of strong ($W_{\lambda2796}>
0.6$\AA) \mgii systems with associated \feii$\lambda2600$ absorption
such that $1 < W_{\lambda2796}/W_{\lambda2600} < 2$ are confirmed to
be \emph{bona-fide} DLAs by subsequent ultraviolet (UV) spectroscopy.  789
of our \mgii systems with $0.84<\zabs<1.3$ satisfy these requirements and we shall refer to
this sample as the ``Mg~{\sc ii}-selected DLAs''.

\subsection{Strong metal lines associated with \caii absorption systems}

In Fig.~\ref{fig:mgiifeii}, the equivalent widths of
\mgii$\lambda2796$, \feii$\lambda2600$ and \mgi$\lambda2853$ in the \caii
absorbers (filled symbols) are compared with those of the \mgii absorbers
(small dots).  In general, \caii systems tend to have strong
associated Mg~{\sc ii}, Fe~{\sc ii} and Mg~{\sc i} lines.  Referring
to the left-hand panel, it can seen that all but two of the \caii
systems fall within the defining criteria of Mg~{\sc ii}-selected DLAs
and their average equivalent width is significantly higher than the
average for the Mg~{\sc ii}-selected DLAs in general.  Turning to the
right-hand panel, there is a hint that the \caii absorbers may have
stronger \mgi for a given equivalent width of Mg~{\sc ii} than
conventional \mgii absorbers, possibly indicating that they tend to
arise in regions of high gas density \citep[][see also Section
\ref{thehighest}]{1974ApJ...188L.107H}.

\section{Selection function and total redshift path}\label{sec:nz}

In order to assess the statistical properties of our \caii absorption
line systems and compare them with those of other classes of
absorbers, it is necessary to quantify the sensitivity of our survey
as a function of \caii equivalent width and wavelength.  We have
addressed this question with Monte Carlo simulations by placing
artificial \caii and \mgii absorption lines in the SDSS quasar spectra
and determining the fraction recovered using the same search
techniques as in the real survey.  The simulations were run for
doublets with ratios of 1:1, 2:1 and 4:3 for Ca~{\sc ii}, and
1:1 and 2:1 for Mg~{\sc ii} and the effect of varying the
line width was investigated.  The general conclusion from these
different trials was that for the parameter ranges considered there
was a variation in detection efficiency of at most 5-10\%.


\begin{figure}
  {\hspace*{-0.6cm} \includegraphics[scale=0.75]{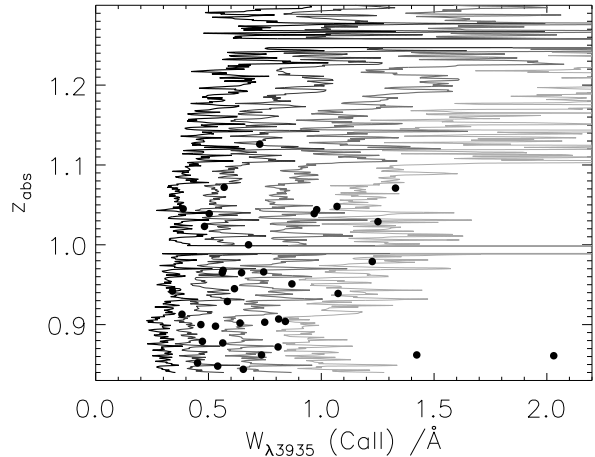}}
  \caption{The probability of detecting a \caii absorption system in
    our survey as a function of absorption redshift and rest
    equivalent width of the stronger member of the \caii doublet,
    calculated from a series of Monte Carlo trials.  From left to
    right, the contours are drawn at probabilities $P_{\rm Ca} = 0.1,
    0.3, 0.7$ and 0.95.  In the example shown here, we assumed a \caii
    doublet ratio $W_{\lambda3935}:W_{\lambda3970} = 4:3$ and that a
    \mgii doublet is detected for each \caii system recovered by our
    search technique, except when blended with the strong sky lines at
    $\lambda5579$ and $\lambda6302$. These sky lines give rise to
    the conspicuous gaps in our detection efficiency at $\zabs \simeq
    0.99$ and 1.25.  Filled circles correspond to the 37 \caii
    absorbers detected in our survey.}
   \label{fig:nz1}
\end{figure}

Fig.~\ref{fig:nz1} shows the probability of detection, $P_{\rm Ca}$,
of a \caii system with doublet ratio $W_{\lambda3935}:W_{\lambda 3970}
= $\,4:3 as a function of redshift $z_{\rm abs}$ and equivalent width
of the stronger member of the doublet, $W_{\lambda3935}$.  In this
example we have assumed that a \mgii doublet is detected for every
artificial \caii system recovered (thus setting $P_{\rm Mg}=1$),
except at redshifts which place the \mgii lines close to strong sky
emission lines.  Values of $z_{\rm abs}$ and $W_{\lambda3935}$ for the
37 \caii systems are overplotted as filled circles.  The probability
of detection drops sharply at redshifts $\zabs > 1.2$, as the \caii
doublet moves beyond 8500\,\AA\ in the observed frame where the
spectra are of poorer quality; the pathlength
available for finding absorbers also decreases with increasing
redshift.  Even so, it is perhaps surprising that \emph{no} \caii
systems have been detected at these redshifts. We test the
significance of this result below.

\begin{figure}
  {\hspace*{-0.3cm} \includegraphics[scale=0.75]{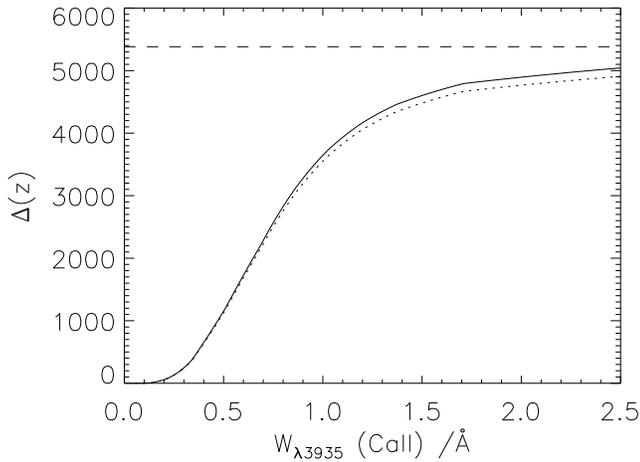}}
  \caption{The total redshift path of our \caii survey calculated from
   Monte Carlo trials. In the two examples shown here, the artificial
   \caii lines used in the simulations had a fixed doublet ratio of
   4:3, and the corresponding \mgii lines were either always detected
   (continuous curve) or assumed to have $W_{\lambda2796} =
   W_{\lambda2803} = 1.0$\,\AA\ (dotted curve).  The horizontal
   long-dash line shows the maximum pathlength available if the
   probability of detection were $P_{\rm detect} = 1$ 
   over the entire redshift path covered by each input quasar
   spectrum.}
  \label{fig:nz2}
\end{figure}

By multiplying the number of quasar spectra in which a \caii absorption
system at a given redshift could be detected, ${\cal N}_{\rm
quasar}$~($z_{\rm abs}$), by the probability of detection given its
values of $\zabs$, $W_{\lambda3935}$ and $W_{\lambda2796}$, we
obtain an estimate of the \emph{effective} number of sightlines over which we
could expect to find such a system:
\be 
{\cal N}_{\rm eff} (\zabs, W_{Ca}, W_{Mg}) = {\cal N}_{\rm quasar}(\zabs)
\times P_{\rm detect} 
\ee 
where
\be
P_{\rm detect} = 
P_{\rm Ca}(W_{\rm Ca},\zabs) \times P_{\rm Mg}(W_{\rm Mg},\zabs)
\ee
and the notation for the rest frame equivalent widths of \caii$\lambda
3935$ and \mgii$\lambda2796$ has been shortened for convenience.

Integrating over redshift gives the total redshift path covered by our
survey as a function of rest frame equivalent width:
\be 
\Delta Z(W_{\rm Ca}, W_{\rm Mg}) =
\int_{z_{\rm abs,min}}^{z_{\rm abs,max}}
{\cal N}_{\rm eff} \, (\zabs,W_{\rm Ca},W_{\rm Mg})\, d\,\zabs.
\label{eq:dz}
\ee 
This function is plotted in Fig.~\ref{fig:nz2} for two
cases: one calculated assuming $P_{\rm Mg} = 1$ throughout (except
when the \mgii lines are blended with sky lines) and the other
adopting $W_{\lambda2796} = 1.0$\,\AA\ (a conservative lower limit for
the majority of \caii absorbers) and a \mgii doublet ratio of 1:1.  The difference
between the two cases is minimal.

It was noted above that no absorption systems are found between
$1.2<\zabs<1.3$ and we are now in a position to calculate the
significance of this result.  The total pathlength available for
finding absorbers in this redshift range is only 0.064 of that
available over the entire $0.84<\zabs<1.3$ range. By adopting a
binomial distribution with 37 trials and P($\zabs>1.2$) = 0.064, the
probability of detecting 0/37 absorbers in this redshift range is
0.086, i.e. the significance of the lack of detection is $\sim 90$\%.
Perhaps with future SDSS catalogue releases it will be possible to
confirm that the effect is indeed a low significance statistical
fluctuation.

\begin{figure*}
  \hspace*{-0.5cm}
  \begin{center}
  \begin{minipage}{10.5cm}
    \includegraphics[scale=0.475]{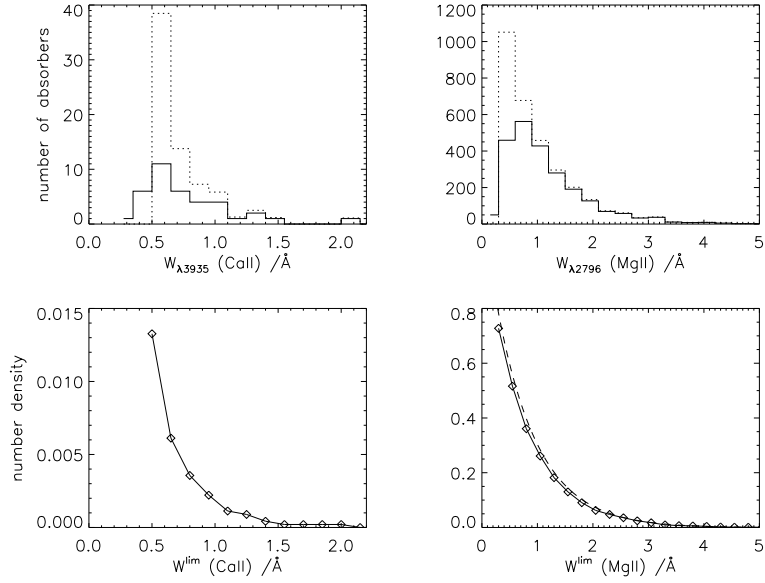}
  \end{minipage}
  \end{center}
  \caption{\emph{Top two panels}: Equivalent width distributions of
    \caii and \mgii absorption systems before (continuous histograms)
    and after (dashed histograms) completeness corrections.  No
    corrections are shown for the smallest equivalent width bins where
    the statistics are too poor for reliable estimates.  \emph{Bottom
    two panels}: Number density of absorbers per unit redshift
    interval as a function of minimum equivalent width limit,$n(W^{\rm
    lim})$.  The values we deduce for the \mgii systems are in good
    agreement with the fit to the distribution of SDSS EDR \mgii
    systems with $0.871 <z_{\rm abs} < 1.311$ (dashed line) reported
    by \citet{2005ApJ...628..637N}.}
  \label{fig:nz3}
\end{figure*}

By combining the observed equivalent width distribution with
Eq.~\ref{eq:dz}, an estimate can be made of the intrinsic
equivalent width distribution of the \caii absorbers.  This is shown
alongside the \mgii equivalent width distribution in the two top
panels of Fig.~\ref{fig:nz3}.  We only show the completeness
corrections above 0.5\,\AA \ and 0.3\,\AA \ for the \caii and \mgii
samples respectively; below these values the corrections become large
and uncertain due to the small number of systems.

Finally, a statistic commonly used in absorption line surveys is the
number of systems per unit redshift with rest equivalent width greater
than some threshold value: $n(W^{\rm lim})$.  While in general
$n(W^{\rm lim})$ is a function of redshift, we assume negligible
redshift dependence here since we are dealing with only a small
redshift interval.  By setting the minimum equivalent width of
\mgii$\lambda2796$ to be 0.6\,\AA, smaller than measured for any of
the \caii systems in our survey, we can remove the joint dependency in
the cumulative probability function:

\be
n(W^{\rm lim}_{\rm Ca},W^{\rm lim}_{\rm Mg} = 0.6) =
\sum_{i}
\frac{1}{\Delta Z_i(W_{\rm Ca}\ge W^{\rm lim}_{\rm Ca})} 
\label{eq:nz}
\ee
where the sum is over all absorbers in the sample with
$W_{\lambda3935}$ greater than the limit.  The function $n(W^{\rm
lim}_{\rm Ca},W^{\rm lim}_{\rm Mg} = 0.6)$ is shown in the bottom
left-hand panel of Fig.~\ref{fig:nz3}; we find that for $W^{\rm
lim}_{\rm Ca~II~\lambda3935} = 0.5$\,\AA, the \caii systems have a
number density per unit redshift $n(z) = 0.013$ (at a mean redshift
$\langle z_{\rm abs} \rangle = 0.95$). 

We also calculated $n(W^{\rm lim})$ for the \mgii systems (Section
\ref{sec:mgii}); the results are shown in the bottom right-hand panel
of Fig.~\ref{fig:nz3}.  Our estimates are in good agreement with those
determined by \citet{2005ApJ...628..637N} from the SDSS EDR (dashed
line) over a very similar redshift range.

\section{Element ratios in \caii absorbers}\label{sec:metals}
  
\begin{figure*}
  \hspace*{-0.5cm}
  \begin{minipage}{\textwidth}
    \includegraphics[scale=0.85]{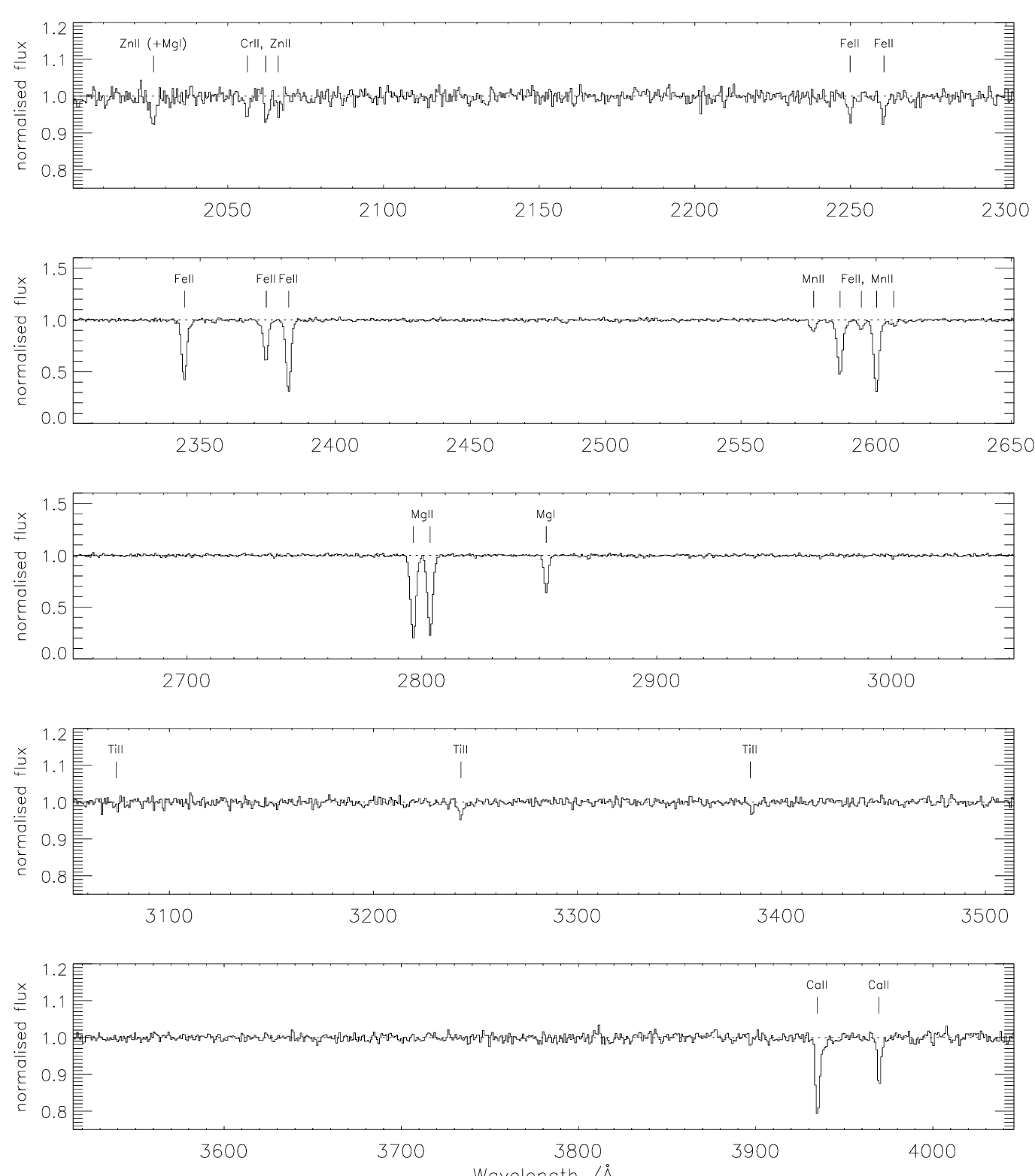}
  \end{minipage}
  \vspace{0.25cm}
  \caption{Composite spectrum of the 27 \caii absorption systems with
    $\langle \zabs \rangle =  0.98$; transitions listed  in Table~\ref{tab:dipso} are
    indicated. Note that two different $y$-axis scalings are used
    for the weak and strong lines respectively.}
  \label{fig:vpfitca}
\end{figure*}

\subsection{Composite spectra}\label{sec:comp1}

In this section we deduce the column densities and relative abundances
of different ions associated with the \caii absorbers whose absorption
lines are covered in the SDSS spectra.  The chemical composition of
the \caii systems should help clarify their relationship to other
classes of quasar absorption line systems, particularly DLAs.  We are
especially interested in investigating the depletion pattern of
refractory elements for comparison with the reddening results reported
in Paper I and re-examined in Section \ref{sec:redden} below.  The
unsaturated absorption lines which are required for precise abundance
analyses are generally too weak to be detected and measured in the
individual SDSS quasar spectra at our disposal.  We circumvented this
difficulty by adding together, in the absorber rest frame, the spectra
of quasars with \caii systems to form composites of sufficient SNR.
This technique has also been applied to absorption systems in SDSS
spectra by \citet{2003ApJ...595L...5N}.

Out of the 37 quasar spectra in our \caii catalogue, 27 encompass the
redshifted wavelengths of the \znii$\lambda\lambda2026, 2062$ doublet
in the \caii absorption system.  Given the diagnostic importance of
this ion \citep[e.g][]{1990ApJ...348...48P}, we limited the abundance
analysis to this subset of 27 \caii absorbers.  We considered three
composite spectra: one containing all 27 absorbers, and two further
subsets each containing about half of the spectra (13 or 14) separated
at the median value of the equivalent width of the \caii$\lambda3935$
line, $W_{\lambda3935} = 0.68$\,\AA.  We refer to these three samples
respectively as `All', `High-$W_{\lambda 3935}$', and
`Low-$W_{\lambda3935}$'.

To create the composite spectra, the individual quasar spectra were
shifted to the absorber rest frame without rebinning\footnote{The SDSS
spectra are binned in $\log\lambda$ with a pixel size of 0.0001, thus
on blueshifting, rebinning of each spectrum is not necessary.}.  Large
scale variations in each spectrum were removed by employing first a
spline continuum fit and then a sliding median filter of 45 pixels to
the residual, with strong absorption lines masked. The spectra were
subsequently combined into a composite using an error-weighted
arithmetic mean.  Alternative methods for both flattening the
individual spectra and combining them were explored; in all cases we
found the final result to be insensitive to the precise methods used.
A second normalisation was carried out on portions of the composite
spectra that included absorption lines of interest by spline-fitting
of the continuum, and the rest frame equivalent widths of the lines
were then measured.  Table~\ref{tab:dipso} summarises the results, and
Fig.~\ref{fig:vpfitca} shows the normalised composite spectrum formed
by coadding all 27 \caii absorbers.

\begin{table*}
  \begin{center}
  \begin{minipage}{12cm}
    \caption{\small \label{tab:dipso} Rest frame equivalent widths 
    of metal lines measured in the three \caii absorber
    composites.}
  \begin{tabular}{ccrrrr} \hline\hline
    &&&\multicolumn{3}{c}{ $W$ (m\AA)}\\\cline{4-6}
    \rule [-3mm]{0mm}{8mm} Ion & Wavelength (\AA) 
    & $f$-value\footnote{Rest frame wavelengths and $f$-values from \citet{2003ApJS..149..205M}} 
    & All &
    High-$W_{\lambda3935}$ & Low-$W_{\lambda3935}$\\ \hline
    Ca~{\sc ii} & 3934.775 & 0.6267 &  725$\pm$ 30 & 1011$\pm$ 45 &  489$\pm$ 24\\
    Ca~{\sc ii} & 3969.590 & 0.3116 &  352$\pm$ 29 &  549$\pm$ 44 &  371$\pm$ 29\\
    Ti~{\sc ii} & 3242.918 & 0.232 &  124$\pm$ 14 &  120$\pm$ 22 &   92$\pm$ 19\\
    Ti~{\sc ii} & 3384.730 & 0.358 &   76$\pm$ 15 &    \ldots &   87$\pm$ 18\\
    Mg~{\sc i} & 2852.963 &  1.83 &  782$\pm$ 13 &  928$\pm$ 30 &  717$\pm$ 20\\
    Mg~{\sc ii}& 2796.354 & 0.615 & 2256$\pm$ 13 & 2472$\pm$ 28 & 2191$\pm$ 22\\
    Mg~{\sc ii} & 2803.531 &     0.306 & 2106$\pm$ 13 & 2238$\pm$ 28 & 2067$\pm$ 22\\
    Fe~{\sc ii} & 2600.173 &     0.239 & 1722$\pm$ 15 & 1799$\pm$ 27 & 1598$\pm$ 18\\
    Fe~{\sc ii}  & 2586.650 &   0.0691 & 1357$\pm$ 17 & 1351$\pm$ 26 & 1268$\pm$ 19\\
    Mn~{\sc ii}  & 2606.462 &    0.198 &  147$\pm$ 12 &  122$\pm$ 21 &  100$\pm$ 17\\
    Mn~{\sc ii} & 2594.499 &     0.280 &  213$\pm$ 14 &  231$\pm$ 21 &  146$\pm$ 17\\
    Mn~{\sc ii} & 2576.877 &     0.361 &  277$\pm$ 14 &  395$\pm$ 29 &  227$\pm$ 17\\
    Fe~{\sc ii}  & 2382.765 &   0.320 & 1582$\pm$ 16 & 1606$\pm$ 28 & 1554$\pm$ 19\\
    Fe~{\sc ii}  & 2374.461 &   0.0313 &  929$\pm$ 16 & 1014$\pm$ 30 &  891$\pm$ 20\\
    Fe~{\sc ii}  & 2344.214 &    0.114 & 1325$\pm$ 16 & 1360$\pm$ 29 & 1385$\pm$ 24\\
    Fe~{\sc ii}  & 2260.781 &   0.00244 &  128$\pm$ 17 &  129$\pm$ 30 &  135$\pm$ 20\\
    Fe~{\sc ii}  & 2249.877 &    0.00182 &  109$\pm$ 15 &   95$\pm$ 27 &   87$\pm$ 17\\
    Cr~{\sc ii} & 2066.164 &         0.0512 &   46$\pm$ 12 &   80$\pm$ 16 &   84$\pm$ 17\\
    Cr~{\sc ii}\,(+Zn~{\sc ii}) & 2062.236 & 0.0759 (0.246) &  114$\pm$ 15 &  195$\pm$ 22 &  111$\pm$ 17\\
    Cr~{\sc ii} & 2056.257 &          0.103 &   92$\pm$ 18 &   90$\pm$ 25 &  108$\pm$ 19\\
    Zn~{\sc ii}\,(+Mg~{\sc i}) & 2026.137 &  0.501 (0.113) &  174$\pm$ 19 &  267$\pm$ 26 &  117$\pm$ 19\\ \hline

  \end{tabular}
  \vspace{-0.4cm}
  \end{minipage}
  \end{center}
\end{table*} 

\subsection{Element abundances and dust depletions}\label{sec:vpfit}

Inspection of Table~\ref{tab:dipso} and Fig.~\ref{fig:vpfitca} shows that our
composite spectra cover a variety of transitions from the elements Mg, Ca, Ti,
Cr, Mn, Fe and Zn.  To avoid potential problems with line saturation, we limited
our analysis to lines with apparent optical depth less than $\sim 0.1$, which
are most likely to fall on the linear part of the curve of growth.  Unless
the distribution of equivalent widths of these lines among the 27 absorption
systems is highly non-uniform---something which we can exclude from inspection
of the individual spectra---the column densities we deduce from these weak
transitions should be representative of the typical values for the sample as a
whole.  In any case, any non-linearities which may be affecting the absorption
lines in the composite spectra would lead us to \emph{under}estimate the
corresponding column densities.

\begin{table*}
  \begin{center}
     \begin{minipage}{14cm}
    \caption{\label{tab:vpfit} \small Ion column densities and element abundances
    relative to Zn. }
   \begin{tabular}{lccccrrr} \hline\hline
      &&\multicolumn{3}{c}{$\log N$} &
       \multicolumn{3}{c}{[X/Zn]}\\\cline{3-5}\cline{6-8} \rule
       [-3mm]{0mm}{8mm} Ion & Solar\footnote{Solar abundance relative to hydrogen,
       in the usual logarithmic scale with H at 12.00 \citep{2003ApJ...591.1220L}.}
        & All & 
       High-$W_{\lambda3935}$ & Low-$W_{\lambda3935}$ & All & 
       High-$W_{\lambda3935}$ & Low-$W_{\lambda3935}$ \\ \hline

       \znii &  4.63 & 12.82$\pm$0.065 & 13.16$\pm$0.054 & 12.59$\pm$0.110 &   0.00 &   0.00 &   0.00\\
       \crii &  5.65 & 13.39$\pm$0.068 & 13.49$\pm$0.093 & 13.42$\pm$0.065 &  $-$0.45 &  $-$0.69 &  $-$0.19\\
       \feii &  7.47 & 15.09$\pm$0.045 & 15.02$\pm$0.086 & 15.09$\pm$0.048 &  $-$0.56 &  $-$0.98 & $-$0.34\\
       \tiii &  4.92 & 12.48$\pm$0.083 & 12.31$\pm$0.123 & 12.45$\pm$0.069 & $-$0.63 &  $-1.14$ &  $-0.43$\\
       \caii &  6.34 & 12.94$\pm$0.017 & 13.10$\pm$0.019 & 12.86$\pm$0.023 &  
       $> -1.59$ \footnote{The values for \caii are lower limits because a significant
       but unknown fraction of Ca is doubly ionised and the line
       ratios of some individual absorbers suggest a degree of saturation.} &  $> -1.77$ &  $> -1.45$\\
       \mnii &  5.50 & 13.09$\pm$0.023 & 13.19$\pm$0.030 & 12.96$\pm$0.032 &  $-0.60$ &  $-0.84$ &  $-0.50$\\ \hline
      \end{tabular}
    \vspace{-0.4cm}
  \end{minipage}
  \end{center}
\end{table*}

Ion column densities, $N$, were derived by fitting all the weak lines
in the composite spectra with the line profile fitting package
VPFIT\footnote{http://www.ast.cam.ac.uk/$\sim$rfc/vpfit.html} adopting
$f$-values and rest wavelengths from the compilation by
\citet{2003ApJS..149..205M} and reproduced in Table~\ref{tab:dipso}.
The profile fitting approach was chosen primarily to ensure internal
consistency in absorption redshift and width among the lines; for such
weak lines the column density is independent of the details of the
line profile and essentially the same values of $N$ would have been
deduced with the analytical expression which relates equivalent width
and column density for lines on the linear part of the curve of
growth:

\be
N = 1.13 \times 10^{20} \cdot \frac{W}{\lambda^2 \, f} \hspace {0.2cm} {\rm cm^{-2}}
\ee
where $W$ and $\lambda$ are both in \AA ngstroms.

Columns 3-5 of Table~\ref{tab:vpfit} list ion column densities in each
of the three composites together with the errors returned by VPFIT.
In all but one case (Ca~{\sc ii}), the ions observed are the major
ionisation stages of the corresponding elements in \hi regions.  Thus,
under the working hypothesis that \caii absorbers with equivalent
width $W_{\lambda3935} \gsim 0.35$\,\AA\ (the lower limit of our sample)
are DLAs (where the gas is mainly neutral and ionisation corrections
are small), we take the ion ratios relative to \znii as measures of
the elements' abundances relative to Zn.  Comparison with the solar
scale [we adopt the set of values proposed by
\citet{2003ApJ...591.1220L}] finally gives the entries of
[X/Zn]\footnote{We use the usual notation whereby ${\rm
[X/Zn]}\,\equiv \log {\rm (X/Zn)}_{\rm obs} - \log {\rm
(X/Zn)}_{\odot}$.}  in the last three columns of
Table~\ref{tab:vpfit}.


 \begin{figure}
  \hspace*{-0.5cm}
  \vspace*{-1.5cm}
  \begin{minipage}{9cm}
  \includegraphics[scale=0.75]{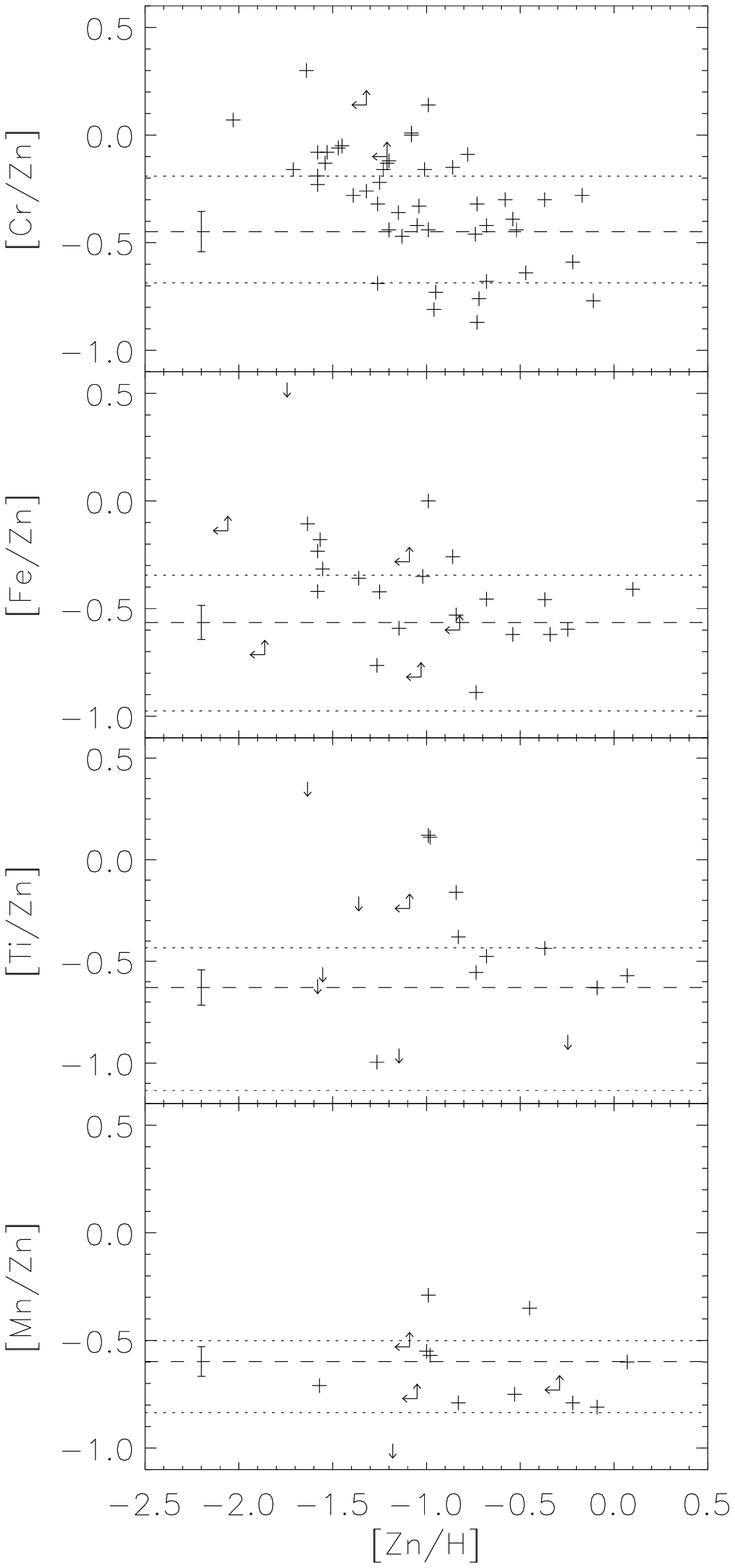}
  \caption{The crosses show the abundances of refractory elements
    relative to Zn (an element which shows little affinity for dust)
    in DLAs from measurements reported in the literature.  The
    horizontal lines are plotted at the values of [X/Zn] in
    Table~\ref{tab:vpfit}: the dashed lines (with error bars) are the
    values for the sample of all 27 \caii absorbers, while the dotted
    lines are for the two subsamples of `High-' (lower dotted line) and
    `Low- $W_{\lambda3935}$' (upper dotted line). The stronger \caii
    absorbers consistently exhibit a higher degree of dust depletion.
    The DLA measurements generally refer to a wider redshift range
    than that of the \caii sample considered here.  The sources of the
    DLA data and corresponding redshift intervals are as follows.  Cr:
    \citet{2005ApJ...618...68K}; \citet{CJA05} ($0.69<\zabs<3.39$).
    Fe: \citet{2001ApJS..137...21P}; \citet{1999ApJ...510..576P};
    \citet{2000ApJ...532...65P} ($0.61 < \zabs <3.39$).  Ti:
    \citet{2002A&A...385..802L}; \citet{2001ApJS..137...21P} ($0.43 <
    \zabs < 2.48$).  Mn: \citet{2002A&A...385..802L}
    ($0.43<\zabs<2.14$). Many of these measurements were obtained from the HIRES DLA
    database at: http://kingpin.ucsd.edu/$\sim$hiresdla/ .}
  \label{fig:dla}
  \end{minipage}
\end{figure}

While we cannot deduce the overall metal abundances of the \caii
absorbers without an independent measure of $N$(H~{\sc i}), we can
still compare their depletion pattern to that commonly encountered in
DLAs.  We have referred the ratios to \znii column density because,
alone among the elements available to us, the depletion level of Zn
onto dust grains is low in the Milky Way ISM, compared to the large
and variable depletions of the refractory Ti, Cr, Mn and Fe relative
to the solar values \citep{1996ARA&A..34..279S}. In reality, intrinsic
differences in the nucleosynthetic history of the elements can lead to
departures from solar relative abundances; dust depletion
subsequently operates on this underlying pattern.  While this is a rich
topic of study \citep[e.g.][]{wolfe05}, it is not necessary to
differentiate between the two effects in the current study as we are
primarily concerned with an overall comparison between the \caii
absorbers and DLAs.

To this end, we have searched the available DLA literature for systems
where the abundances of Zn and at least one of the four refractory
elements considered here have been measured.  After adjusting the
abundances to Lodder's (2003) solar scale, we compare them in
Fig.~\ref{fig:dla} with those of the \caii absorbers.  References to
the original papers are given in the figure caption.  Note that in
general the measurements from the literature refer to a much wider range of redshifts
than probed here.

Fig.~\ref{fig:dla} shows clearly that the element depletions deduced
for the \caii systems are typical of the values encountered in most
DLAs, with Ti, Cr, Mn and Fe on average less abundant than Zn by
factors of 3-4 ($\sim 0.5$-0.6 dex).  Their is a marked distinction
between the two subsamples, with the `High-$W_{\lambda3935}$'
subsample in particular exhibiting some of the most pronounced
depletions measured in DLAs up to now.  It is certainly of
considerable interest that similar levels of [Fe/Zn] depletion are
seen in DLAs with significant molecular hydrogen contents, where dust
is believed to be the catalyst for molecular hydrogen formation
\citep{2003MNRAS.346..209L}.  Most data are available for the Zn and
Cr pair (top panel of Fig.~\ref{fig:dla}), partly for historical
reasons \citep{1990ApJ...348...48P} and partly because, at the
redshifts of most DLAs, the relevant absorption lines are conveniently
located in the optical spectrum. The ratio [Cr/Zn] shows an
approximate correlation with [Zn/H]: the depletion of Cr decreases
with decreasing overall metallicity, although the scatter is
considerable \citep{1997ApJ...478..536P, CJA05}.  By using this
correlation as a rough indicator of metallicity for the \caii
absorbers, a range between $\sim 1/30$ and $\sim 1/3$ of solar is
suggested ([Zn/H] between $-1.5$ and $-0.5$).


\begin{figure}
  \includegraphics[scale=0.65]{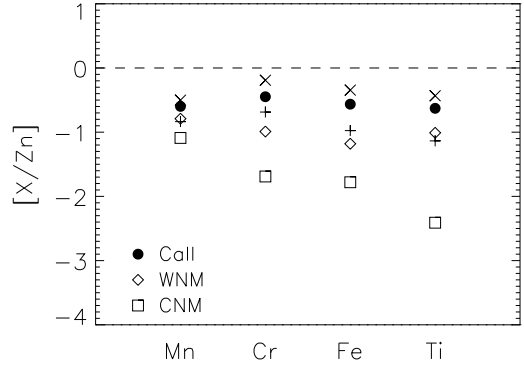}
  \caption{Abundances relative to Zn of refractory elements in the \caii
    absorbers (indicated by crosses), and in the warm (diamonds)
    and cold (squares) neutral interstellar medium of the Milky Way
    \citep[from the compilation by][]{1999ApJS..124..465W}.
    Three sets of crosses are shown, respectively for all \caii
    systems (black/middle crosses), and for the `Low-'
    (red/upper) and `High-$W_{\lambda3935}$' (blue/lower) 
    subsamples.
    Elements are ordered
    by increasing condensation temperature \citep{1996ARA&A..34..279S}
    and corrected to the solar reference values used in this paper.}  \label{fig:mw}
\end{figure}

Fig.~\ref{fig:mw} presents the depletions relative to Zn of the \caii
absorbers compared to typical diffuse clouds in the solar
neighbourhood. The overall level of depletion is lower in the \caii
absorbers, similar to results for DLAs which is cited as being
consistent with their lower metallicities \citep{2004A&A...421..479V}.
However, we note that the depletions deduced for the
`High-$W_{\lambda3935}$' subsample approach or exceed those typical of
the warm neutral medium of the Milky Way.

\section{Reddening due to dust}\label{sec:redden}

An alternative method of estimating the dust content of quasar
absorbers is to consider the reddening effect of the dust on the SED
of the background quasar.  This is an intrinsically statistical
approach, because we do not have an \emph{a priori} knowledge of the
unobscured SED of an individual quasar, but rather we have to rely on
the typical SED of an ensemble of quasars with properties broadly
similar to those being tested for extinction.  For this reason, the
availability of large control samples of quasar spectra afforded by
the SDSS has allowed such tests to be carried out to unprecedented
levels of sensitivity.    \citet{2004MNRAS.354L..31M} concluded
that high redshift ($z_{\rm abs} \sim 3$) DLAs in the SDSS DR2 have an
average \EBV $\lsim 0.02$ based on the lack of any reddening signal.
On the other hand, in Paper I we reported an average \EBV $= 0.06$ due
to the 31 \caii absorbers at $\langle z_{\rm abs} \rangle = 0.95$ in
SDSS DR3.  We also found that splitting that \caii sample at
$W_{\lambda3935} = 0.7$\,\AA\ resulted in \EBV$ = 0.099$ and $0.025$ for
the high and low equivalent width subsets respectively---a trend which
we now see to be consistent with that of the element depletions
discussed above.

In this paper we extend the analysis of Paper I in two ways.  First,
we perform our reddening analysis on the expanded sample of 37 \caii
absorbers, which includes the six newly discovered systems from DR4,
and on the 27 \caii systems analysed in the previous section to allow
direct comparisons of \EBV\ values with Zn column densities and
depletion patterns.  Second, we repeat the analysis on the Mg~{\sc
  ii}-selected DLAs [those \mgii absorption line systems fulfilling the
criteria of \citet{astro-ph/0505479}, Section \ref{sec:mgii}] for
comparison with the \caii absorbers and with the confirmed DLAs at $z
\sim 3$ considered by \citet{2004MNRAS.354L..31M}.

\subsection{Quasar reference spectra and measurement of average reddening}\label{sec:qsorefs}

We look for a reddening signal in the spectra of quasars with an
intervening absorber by comparing their SEDs with that of a control
spectrum: a composite constructed by averaging many spectra of quasars
without those absorbers.  The large number of quasars in the SDSS
allows us to create control composites in small steps of quasar emission redshift
$\zem$.  Details of the method have been described in Paper I and are
only summarised here.  Using the same sample of SDSS DR3 quasars which
were searched for \caii absorption systems (Section \ref{sec:sample}),
but removing all spectra with identified \caii or \mgii absorption
lines, we constructed composite spectra in bins of $\Delta z_{\rm em}
= 0.1$ staggered by $z=0.05$.  Each spectrum was first corrected for
Galactic reddening using the quoted extinction in the SDSS photometric
catalogue and the Galactic extinction curve from
\citet{1989ApJ...345..245C} as extended by
\citet{1994ApJ...422..158O}.  The spectra were then shifted to the
quasar rest frame without rebinning, being careful to allow for the
flux/\AA\ term in the SDSS spectra, scaled by the median flux over a
common rest wavelength range for the particular redshift bin (avoiding
quasar emission lines, strong sky lines and bad pixels), and finally
combined into a composite spectrum using a simple arithmetic mean.
Different methods of combining spectra were tested, and we concluded
that the details of the procedure made no significant difference to
the final composites.

The procedure adopted here differs from that of Paper~I in two small
details.  First, when constructing the reference spectra we excluded
all quasars with Mg~{\sc ii} absorbers (as well as the few quasars
with \caii systems), since we will be testing for reddening caused by
both classes of absorption line system.  Second, we dropped the
correction for potential dependence of the quasar SED on apparent
magnitude, since any effect was found to be insignificant.

Turning now to the spectra of quasars \emph{with} absorbers: each one
of these was shifted to the rest frame of the quasar, scaled by the
median flux as above, and then divided by the reference composite
quasar spectrum closest in redshift.  The resulting ``difference''
spectrum, which contains any reddening signal, was then shifted to the
rest frame of the absorber, $\zabs$, before being combined with others
to form the composite spectra to be analysed for reddening. This
combination was performed using an arithmetic mean; a popular
alternative for this type of analysis is the geometric mean. Employing
a geometric rather than arithmetic mean results in reddening values
10-20\% higher due to enhanced weighting given to the difference
spectra with the largest slopes towards the blue end. In the
composites studied in this paper the differences are insignificant, in
almost all cases they are within the $1\sigma$ errors (68th percentile, see
\ref{subsec:mc}).

We constructed several different composite spectra for both the \caii
absorbers and Mg~{\sc ii}-selected DLAs.  For the
former, we produced three composites equivalent to the `All',
`High-$W_{\lambda3935}$', and `Low-$W_{\lambda3935}$' samples of the
previous section but containing all 37 absorbers. We also produced
reddening composites for the three samples containing absorbers with
available Zn~{\sc ii} for direct comparison with the measured element
column densities.  For the Mg~{\sc ii}-selected DLAs, we considered
the full sample of 789 quasars and three subsamples with increasing
absorption line strength. The subsamples were defined by drawing
diagonal lines across Fig.~\ref{fig:mgiifeii} from ($W_{\lambda2796},
\, W_{\lambda 2600}) = (m,0) $ to ($W_{\lambda2796}, \,
W_{\lambda2600}) = (0,m)$, where $m = 1.5, 2$ and $3$\,\AA, and taking
only the systems that lie above each line and are within the
\citet{astro-ph/0505479} limits.

The reddening of each composite was measured by fitting extinction
curves of three forms, appropriate for dust in the Milky Way, the
Large Magellanic Cloud (LMC) or the Small Magellanic Cloud (SMC) to
find the best fitting value of \EBV.  We used the extinction curve
tabulations of \citet{1989ApJ...345..245C}, as extended by
\citet{1994ApJ...422..158O}, for the Milky Way and of
\citet{1992ApJ...395..130P} for the Magellanic Clouds and a
total-to-selective extinction ratio $R_V = 3.1$ was assumed
throughout.

\subsection{Assessment of minimum reddening detectable}\label{subsec:mc}

\begin{figure}
  \includegraphics[scale=0.6]{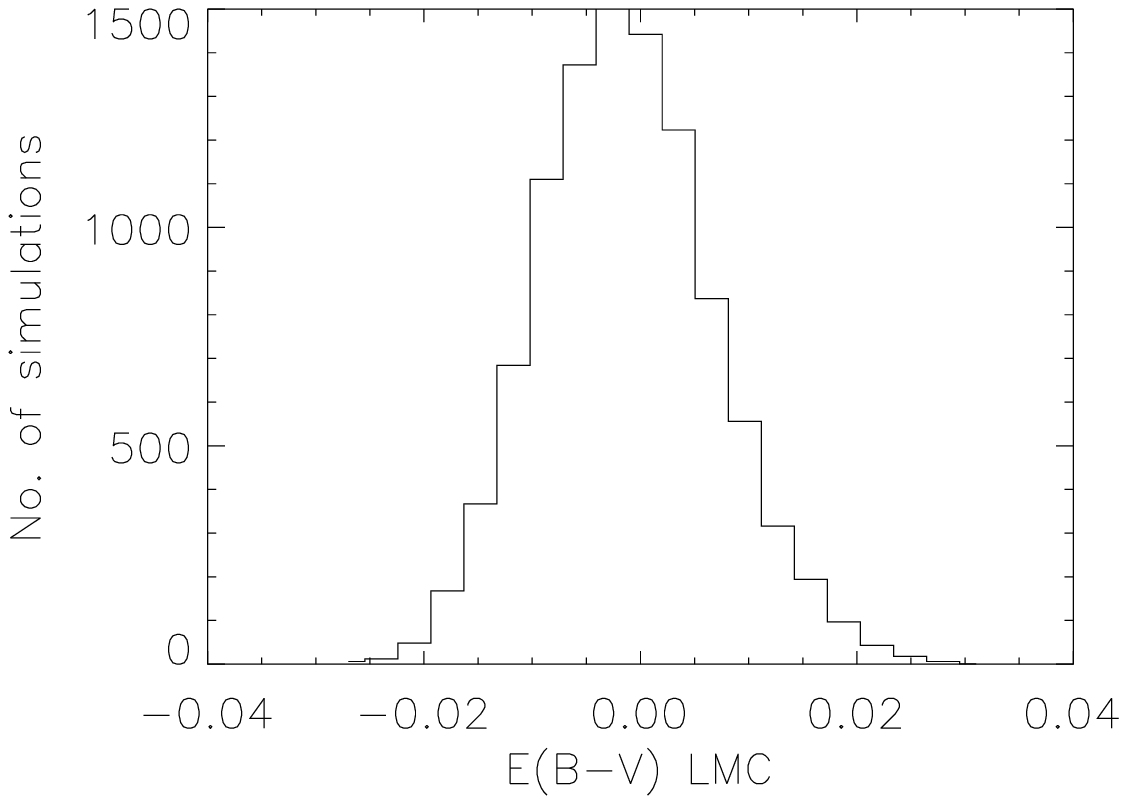}
  \includegraphics[scale=0.6]{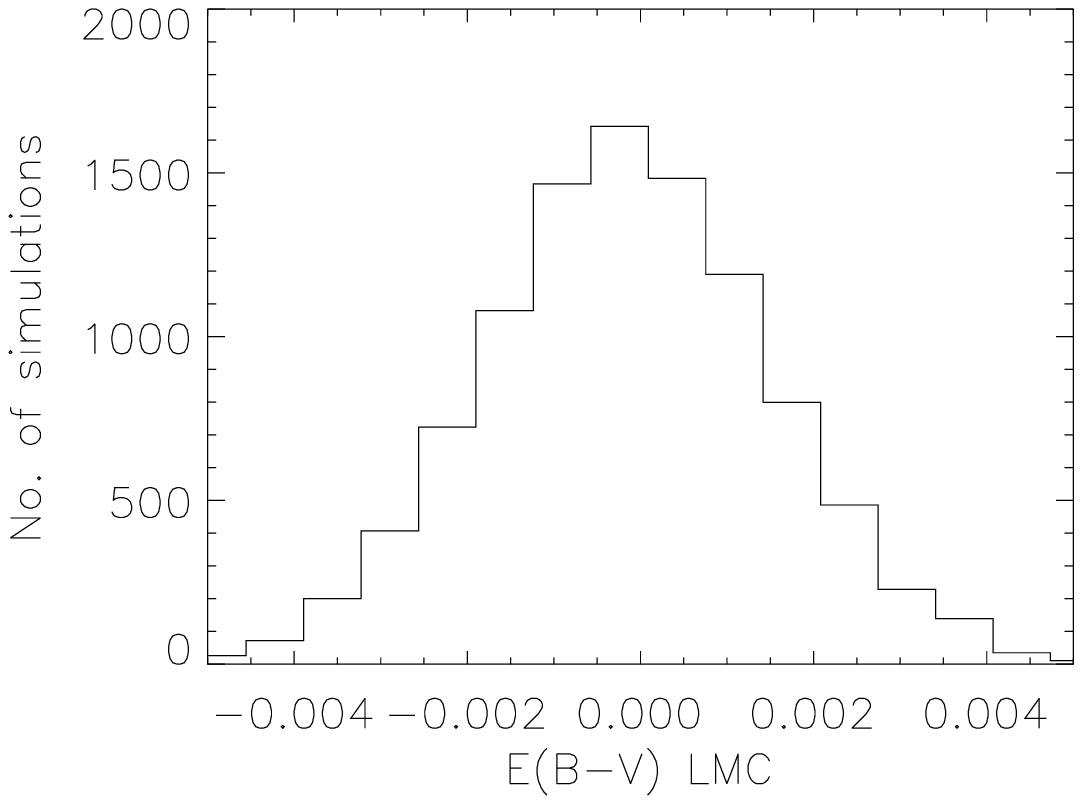}
  \caption{The distribution of apparent values of \EBV\
   produced by Monte Carlo simulations of 10\,000 sets of 37 (top) and
   789 (bottom) randomly chosen quasar spectra when shifted randomly
   to rest frames corresponding to the range in redshift of our
   absorption systems. See text for additional details.  The range of
   \EBV\ covered by the central 68\% of simulations
   provides an estimate of the error on the reddening measured in the
   absorber composite spectra.}
  \label{fig:monte}
\end{figure}


\begin{figure*}
  \vspace*{-0.5cm}
  \hspace*{-0.5cm}
  \begin{center}
   \begin{minipage}{14.75cm}
   \includegraphics[scale=0.7]{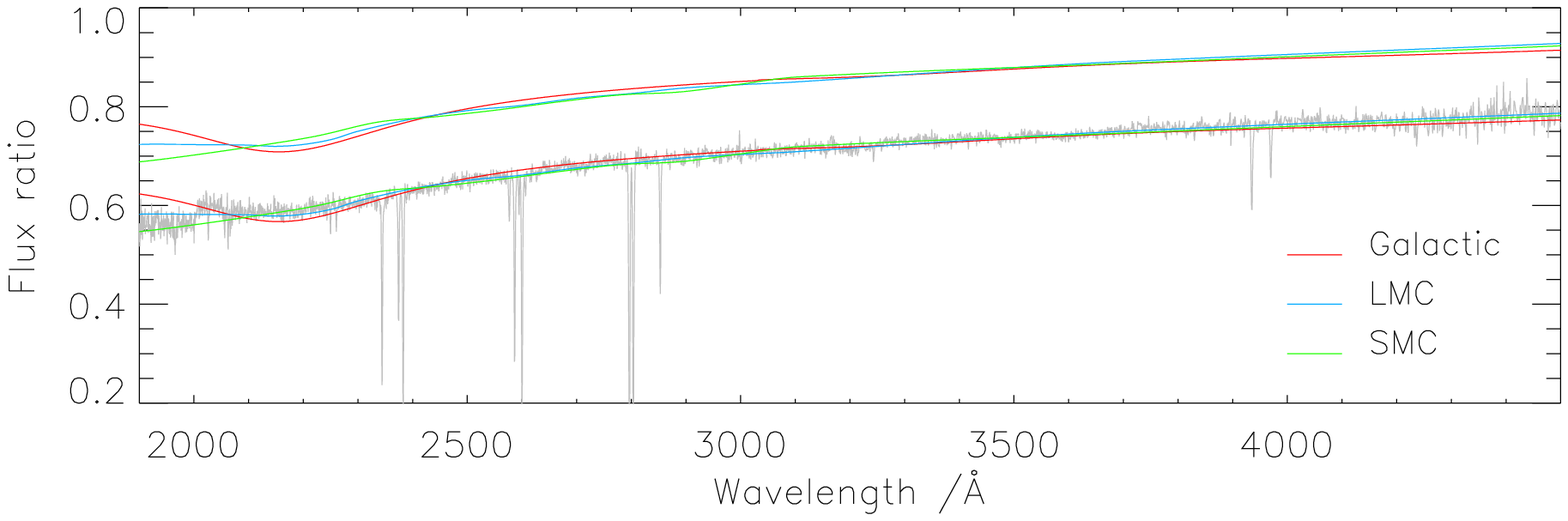}
    \includegraphics[scale=0.7]{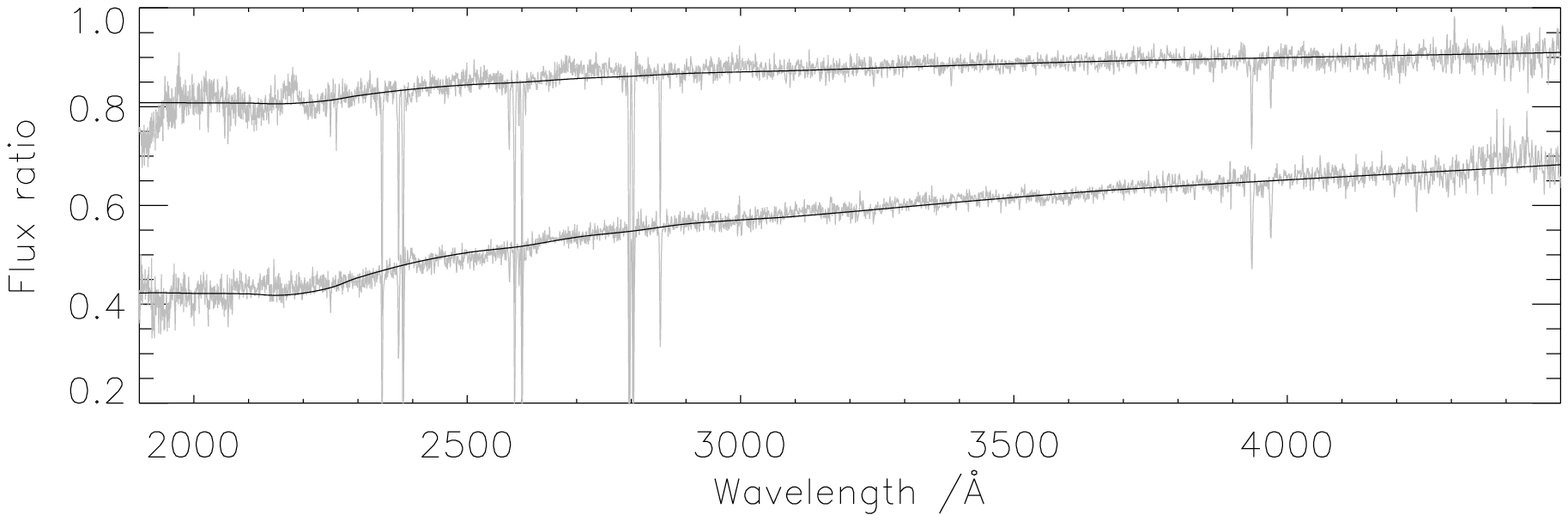}
  \end{minipage}
  \end{center}
  \caption{\emph{Top}: Composite spectrum of the 37 quasars with \caii
    absorbers after an estimate of the unabsorbed quasar SED has been
    divided out and each spectrum has been shifted to the rest frame
    of the absorber before being combined into the composite. The
    resulting spectrum is fitted with extinction curves
    appropriate to dust in the Milky Way and the Magellanic Clouds (as
    indicated) to deduce the values of colour excess, \EBV, listed in
    Table~\ref{tab:EBV}.  The fits to the SED are shown superposed on
    the spectrum, and are also plotted above it, for clarity.
    The spectrum and reddening curve normalisation is calculated
    from the derived \EBV\ values and does not play a role during the
    fitting of the dust reddening curves.
    \emph{Bottom}: As for the upper panel, but with the \caii sample
    now split into the `High-' and `Low-$W_{\lambda3935}$' subsamples.
    The former shows the higher extinction (lower spectrum in the
    bottom panel). Only the LMC extinction curve fits are shown. See
    the online version for a colour figure.}
   \label{fig:cared}
\end{figure*}

Before discussing the results of the reddening analysis, we need to
assess the possibility of any systematic effects which may produce a
spurious reddening signal.  We investigated this with a series of
Monte Carlo simulations in which we constructed artificial composites
by drawing quasar spectra at random from the reference quasar sample
without \caii and \mgii absorption systems.  The selection
probability of each spectrum was modified by the redshift path $\Delta
z_{\rm quasar}$ over which we searched the spectrum for intervening
absorbers:
\begin{eqnarray}\label{eq:deltaz}
\Delta z_{\rm quasar} &=&  min[1.3,\zem-0.01] \nonumber\\ 
&-&  max[0.84,1250(1+\zem)/2175 - 1]
\end{eqnarray}
where $min$ and $max$ indicate the minimum and maximum of the values
enclosed in the brackets.  On selection, each quasar was assigned a
random absorption redshift (within the range over which we searched
for real absorbers), and shifted to the rest frame of this
``absorber''.  The process was repeated until we had assembled
artificial samples of the required size (e.g. 37 for the entire sample
of \caii absorbers).

Composite spectra were then constructed for these artificial absorber
samples and analysed for reddening as described above, assuming an LMC
extinction curve, and the whole process was repeated 10\,000 times.
The final product of this exercise are the distributions of apparent
\EBV\ values which arise simply from variations in the SED of quasars.
The two distributions from the simulations designed to mimic the
entire \caii and Mg~{\sc ii}-selected DLA samples, are reproduced in
Fig.~\ref{fig:monte}. The distributions have a mean \EBV\ close to zero
and possess a small dispersion, extending to only a few thousandths of
a magnitude for the large \mgii sample and to a few hundredths of a
magnitude for the \caii absorbers.  In none of the 10\,000 simulations
performed do we find values of \EBV\ greater than 0.031 and 0.0065 for
the 37 Ca~{\sc ii} absorbers and 789 Mg~{\sc ii}-selected DLAs
respectively. By taking the 68th percentiles of these distributions,
we obtain estimates of the errors on the reddening measured in the
absorber samples for use in the analysis of the results.


  \begin{figure*}
  \hspace*{-0.5cm}
  \begin{center}
   \begin{minipage}{14.75cm}
    \includegraphics[scale=0.7]{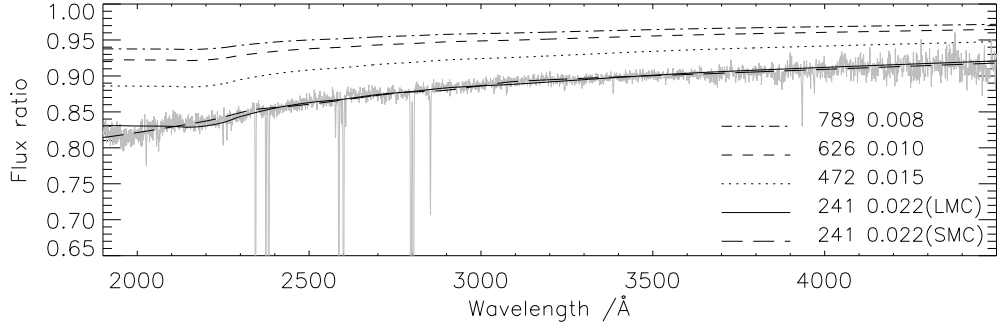}
  \end{minipage}
  \end{center}
  \caption{As for Fig.~\ref{fig:cared} but for Mg~{\sc ii}-selected
  DLA candidates. The curves shown refer to different subsamples of
  Mg~{\sc ii}-selected DLAs ordered by increasing values of
  $W_{\lambda2796}$ and $W_{\lambda2600}$, as described in the
  text. The full sample includes 789 quasar spectra and shows an
  average \EBV$ = 0.008$ (for an LMC extinction law), as indicated in
  the lower right-hand corner of the figure.  Corresponding values for
  the other subsamples are also given.  The composite spectrum of the
  subsample of 241 Mg~{\sc ii}-selected DLAs with the highest
  equivalent widths is shown in grey.  Superposed on the spectrum are
  the best-fitting LMC and SMC extinction curves; both give \EBV$ =
  0.022$. The spectrum and reddening curve normalisation is calculated
  from the derived \EBV\ values and does not play a role during the
  fitting of the dust reddening curves.}
   \label{fig:DLAred}
\end{figure*}

\subsection{Results: reddening by \caii and \mgii absorbers}\label{sec:red}

The results of the reddening analysis of the real absorbers are
collected in Table~\ref{tab:EBV} and illustrated in
Figs.~\ref{fig:cared} and \ref{fig:DLAred}.  
The findings of Paper~I in detecting an unequivocal reddening signal
from the \caii absorbers are confirmed by these results.  We further
strengthen the conclusion that among this sample there is considerable
latitude of dust content, with a very pronounced difference between
the `High-' and `Low-$W_{\lambda3935}$' subsamples (see lower panel of
Fig.~\ref{fig:cared}). For both the 'All' and 'High-$W_{\lambda3935}$'
samples reddening values of the observed magnitude are never measured
in the corresponding Monte Carlo simulations, making these results
significant at the $>$99.99\% confidence level. The small \EBV\
deduced for the `Low-$W_{\lambda3935}$' subsample, \EBV$=
0.026$, is only marginally significant ($<3\sigma$) given the results
of the Monte Carlo simulations.

Turning to the Mg~{\sc ii}-selected DLAs, we see from
Fig.~\ref{fig:DLAred} that they exhibit a substantially lower
reddening signal.  For the full sample of 789 absorbers (dot-dash
line) we measure \EBV$ = 0.008$ (LMC) which, while very low, is still
significant at the $>$99.99\% confidence level according to the
results of our Monte Carlo simulations which suggest $1\sigma$ errors
of $0.0016$.

\begin{table}
  \begin{center}
    \caption{\label{tab:EBV} Estimates of the reddening
    \EBV\ introduced by the \caii absorbers in the spectra
    of background quasars. For each reddening curve used, the upper
    values are for the full 37 absorber sample, and the lower values
    are for the subsample of 27 absorbers analysed for element column
    densities. The final two rows give the estimated error for each
    sample derived from Monte Carlo simulations using an LMC dust
    curve (see Section \ref{subsec:mc}).}
      \begin{tabular}{cccc} \hline\hline
      &\multicolumn{3}{c}{\EBV} \\\cline{2-4}
      \rule [-3mm]{0mm}{8mm} Dust law& All & High-$W_{\lambda3935}$ & 
      Low-$W_{\lambda3935}$\\ \hline
      MW  &0.057& 0.092& 0.023\\
          &0.048& 0.082& 0.012\\
      LMC &0.065& 0.103& 0.026\\ 
          &0.055& 0.093& 0.014\\
      SMC &0.066& 0.105& 0.026\\
          &0.055& 0.095& 0.015\\
      err &0.008& 0.011& 0.011 \\
          &0.009& 0.013& 0.014\\
      \hline
     \end{tabular}
  \end{center}
\end{table}

The best-fit LMC extinction curves for the three subsamples split
according to the strength of their strong lines are also shown in
Fig.~\ref{fig:DLAred} with dashed, dotted and continuous lines
corresponding, respectively, to $m=1.5, 2$ and $3$\,\AA\ (see Section
\ref{sec:qsorefs}).  The number of absorbers and value of \EBV$_{\rm
LMC}$ for each subsample are listed in the bottom right-hand corner of
the figure.  For the composite of 241 absorbers with the strongest
\mgii$\lambda2796$ and \feii$\lambda2600$ absorption lines
($m=3$\,\AA; plotted in grey), we also show the best-fit SMC
extinction curve (long dash).  It is evident from
Fig.~\ref{fig:DLAred} that, as is the case for the \caii absorbers,
Mg~{\sc ii}-selected DLA candidates exhibit a clear trend of
increasing dust content with increasing absorption line equivalent
width. Even accounting for the $\sim 57$\% of sub-DLAs in the sample
\citep{astro-ph/0505479}, the average dust content of DLAs at these
redshifts is tiny. However, for the 30\% of absorbers with the highest
absorption line equivalent widths, the average \EBV\ of the DLAs in
the sample may be similar to the average of the \caii
absorbers\footnote{Note that the \caii absorbers were removed from the
\mgii sample and thus do not contribute to these results (Section
\ref{sec:mgii}).}, assuming that sub-DLAs contribute negligible dust
reddening.

From Figs.~\ref{fig:cared} and \ref{fig:DLAred} it can be appreciated
that at these mid-UV wavelengths and low levels of extinction
there are only subtle differences between the three extinction laws
fitted to the data.  Even so, a 2175\,\AA\ bump as strong as that
produced by Galactic dust can probably be excluded for all of the
composites.  The LMC extinction curve, which flattens near 2250\,\AA,
seems to give the best fit to the reddening produced by the \caii
absorbers.  On the other hand, the steeply rising SMC curve appears to
be a better fit to the DLA candidates selected via \mgii and Fe~{\sc
  ii}.  It will be of interest to establish whether these differences
are significant, since they are ultimately related to differences in
the size distribution and composition of the dust particles associated
with different classes of quasar absorbers.

\begin{figure}
  \includegraphics[scale=0.65]{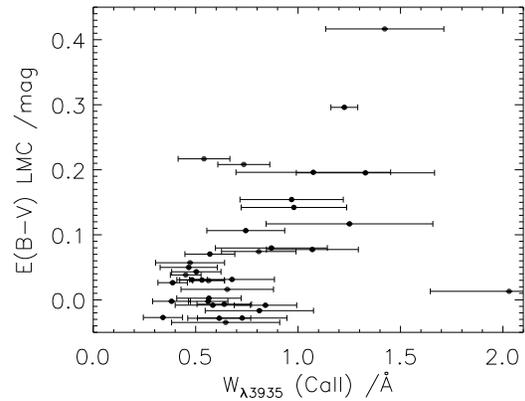}
  \caption{Colour excess due to individual \caii absorbers
  plotted as a function of the equivalent width of \caii$\lambda3935$.}
  \label{fig:EW1}
\end{figure}

Finally, while the results of the reddening analysis are most secure
when applied to ensembles of absorbers, so that variations due to the
intrinsic SEDs of the background quasars are mitigated, it is
nevertheless of interest to examine the absorbers' reddening on a case
by case basis.  The results are also used in Section
\ref{sec:dustbias} for calculating the obscuration bias inherent in
the absorber sample. Each quasar with a \caii system was divided by
the relevant reference spectrum and an LMC extinction curve fitted to
the quotient in the rest frame of the absorber to provide individual
values of \EBV. These values are listed in the last column of
Tables~\ref{tab:1} of the present paper and of Paper~I and in
Fig.~\ref{fig:EW1} we have plotted them against the equivalent width
of \caii$\lambda 3935$.  Despite the scatter, the trend of increasing
reddening with equivalent width highlighted by the analysis of the
'High-' and 'Low-$W_{\lambda3935}$' subsamples can be discerned.


\section{Discussion}

We now draw together the results of the previous sections and discuss
their implications for the use of \caii as a tracer of intermediate
redshift DLAs, the dust-to-metal ratio of the absorbers and compare
their dust contents to high redshift emission-selected galaxies.
Firstly, however, we must address the issue of obscuration bias
against the dustiest, most metal-rich systems.

\subsection{Dust obscuration bias}\label{sec:dustbias}

We begin our discussion with a calculation of the dust obscuration 
bias against the \caii systems, necessary for comparison of their true number
density to that of DLAs. We further investigate the attenuation 
which a similar sample of objects would cause at higher redshifts.

There has been much discussion in the literature concerning possible
selection bias in magnitude-limited quasar samples against dusty,
metal-rich DLAs at high redshift because of the obscuration they would
cause to the background quasar light
\citep[e.g.][]{1984ApJ...278....1O,1989ApJ...337....7F,1993ApJ...402..479F}.
By selecting quasars at radio wavelengths, which are unaffected by
dust, the CORALS survey addressed these concerns
\citep{2001A&A...379..393E,2004ApJ...615..118E}.  No significant
obscuration bias was detected over a wide redshift range. However, the
limited statistics of the samples still allow for the possibility that
$\sim$60\% of DLAs may remain undetected in optical surveys at
intermediate redshifts. In another study, \citet{astro-ph/0502137}
used the observed distribution of column densities of \znii to argue
that between 30 and 50\% of systems would be obscured, although their
conclusions are not borne out by the results of \citet{CJA05}.

The number of \caii absorption line systems missing from the quasar sample,
as a result of the dimming of the background quasars by dust
associated with the \caii absorber, can be determined by considering
the total redshift--path available in the quasar survey with and without
intervening absorbers. 

A particular quasar contributes a certain redshift--path, $\Delta z$
(Eq.~\ref{eq:deltaz}) over which an intervening absorber can be
detected. This is determined via the probability of detection,
$P(W,z_{abs},m_i)$, which is a function of equivalent width of the
line to be found ($W$), the absorber redshift and, because fainter
objects have lower SNR spectra, the observed $i$--band magnitude of
the quasar ($m_i$).  Summing over all quasars in the sample and over
all redshifts for which $P > 0$, produces an estimate of the total
absorber redshift--path for the sample for a given equivalent width
limit.

Absorbers with an associated reddening, \EBV, lying in front of a
quasar, cause a corresponding observed--frame $i$--band extinction of
$A_i\,$mag to the quasar light, making the quasar appear fainter than
otherwise would be the case. The redshift--path is now determined via
the probability of detection $P(W,z_{abs},m_i+A_i)$, where $m_i$ is
again the unextinguished observed $i$--band magnitude of the quasar,
but the quasar now appears in the sample with a measured $i$--band
magnitude $=m_i+A_i$.

For quasars with magnitudes close to the faint magnitude limit of the
sample, the extinction $A_i$ may take the quasar below the faint limit
and $P=0.0$. This reduces the total redshift--path in the sample for which
absorbers causing $A_i\,$mag of extinction to the background quasar
may be identified. Even when the quasar is not lost completely from
the sample the probability $P(W,z_{abs},m_i+A_i) < P(W,z_{abs},m_i)$,
because the SNR of the spectra decreases with increasing
magnitude. The extinction due to the intervening absorber thus makes
it more difficult to find absorption lines of a given equivalent width.

\begin{figure}
  \includegraphics[scale=0.6]{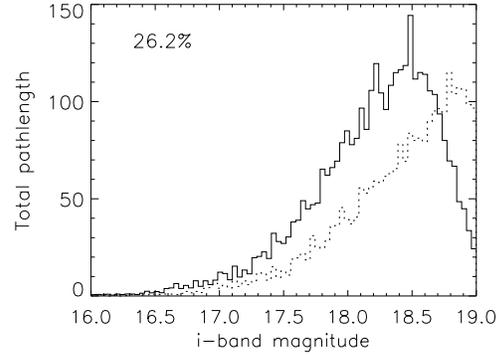}
  \caption{The effect on the total survey redshift--path of dusty
    absorbers. The continuous histogram shows the observed redshift--path
    as a function of $i$-band magnitude of the SDSS quasar sample,
    while the dotted line shows the effect on the distribution were
    each quasar line of sight to be intercepted by an absorber with
    \EBV\,=\,0.065, $\zabs=0.95$ and $W_{\lambda3935}=0.76$ (the mean
    values of the 37 \caii absorbers). See the text for further
    discussion of this figure.  The value in the top left is the
    overall percentage of redshift--path lost.}
  \label{fig:ext}
\end{figure}

Fig.~\ref{fig:ext} shows the redshift--path of the survey available to
find an absorber with $\zabs=0.95$, \EBV\,=\,0.065 (corresponding to
A$_i=0.30$), and $W_{\lambda3935}=0.95$ compared to the redshift--path
available to find the same absorber if it contained no dust
(continuous histogram).  In order to calculate the true number of
absorbers that would be seen in the absence of any ``dust bias'', the
observed number of such absorbers is multiplied by the factor
$\Delta z_{no-dust}/\Delta z_{dust}$.

Calculation of the total number of absorbers missing requires the
availability of a well--determined \EBV\ distribution.  For large
values of \EBV\ we do not have such a distribution and we choose,
conservatively, to calculate the dust obscuration bias only for \caii
absorbers with \EBV$<0.25$.  Performing the calculation for the 35
absorbers with \EBV\ below this limit, an unbiased number of 58
absorbers is found.  Therefore, the obscuration bias towards \caii
absorbers with \EBV$\lsim0.25$ results in a reduction in the observed
number of systems seen in the quasar sample of $\sim$40\% ($1-35/58$).
The $\sim40\%$ figure is insensitive to binning of the data within
sensible ranges of bin size.  These 58 absorbers would have an average
\EBV$\sim0.1$.  Because we have simply omitted those absorbers with
larger \EBV\ from our calculation, these values provide lower limits
for the total sample.


As we discuss below, the dust and metal content of the \caii absorbers
is consistent with their identification as DLAs, and their unbiased
number density is around 20-30\% that of DLAs at similar redshifts.
In Section \ref{sec:red} we also saw that the majority of intermediate
redshift DLAs (that is those selected from their \mgii and \feii
lines) contain negligible quantities of dust.  Based on these results,
we estimate that dust obscuration in the intermediate redshift DLA
population as a whole will cause 8--12\% of DLAs with \EBV$\lsim0.25$
to be missed from optical magnitude limited surveys.  For this \EBV\
limit the results are well within the limits on obscuration bias for
DLAs placed by the CORALS survey \citep{2004ApJ...615..118E}, although
an increasing fraction of DLAs with dust contents greater than this
limit would be obscured.

Briefly, we note that the reddening effect of intervening dust on the
background quasar SED will affect the selection of quasars for
follow-up spectroscopy from the SDSS photometry
\citep{2001AJ....121.2308R}.  Unfortunately, quantification of this
effect is non-trivial at present.

\begin{figure}
  \includegraphics[scale=0.6]{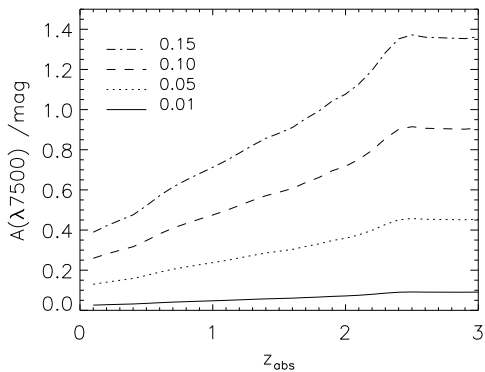}
  \caption{The observed frame ($i$-band) extinction of quasar light
    caused by LMC type dust in an intervening galaxy as a function of $\zabs$. 
    The different curves are for different columns of dust as
    parameterised by the \EBV\ values given in the top left.}
  \label{fig:extz}
\end{figure}

\subsubsection{Bias at higher redshifts}

To illustrate the impact of extinction caused by intervening dust at
higher redshifts on the SDSS quasar sample, Fig.~\ref{fig:extz} shows
the observed frame $i$-band extinction as a function of $\zabs$ caused
by an intervening galaxy with reddening \EBV$_{\rm
LMC}$\,=\,0.01,\,0.05,\,0.1 and 0.15.  It can be seen that an
absorption line system at $\zabs=1$ with \EBV$\sim$0.1 and LMC
type dust causes just under 0.5\,mag of extinction of the quasar
light.  The same system at $\zabs=2.5$ would cause 0.9\,mag of
extinction; the effect of dust obscuration is clearly considerably
greater at the redshifts probed by traditional DLA studies
($\zabs>2.2$ in the SDSS).  

We repeat the dust obscuration analysis for a population of imaginary
\caii absorption systems with the same dust and absorption line
properties as our 35 systems with \EBV$\lsim0.25$, but with redshifts
randomly assigned between 2.2 and 3.  The quasar sample is taken from
the DR3 catalogue of \citet{astro-ph/0503679} with $2.2<\zem<4$ and
$i<20.2$.  Assuming that the identification of DLAs is independent of
quasar magnitude (a reasonable approximation for strong absorption
features such as a damped Lyman-$\alpha$ line rather than weak \caii
absorption), leads to the prediction that $\sim75$\% of the {\it underlying}
population of absorbers with \EBV$\lsim0.25$ would have been obscured
were these observations to have been made at high redshift.  The
calculation strongly suggests that were a subset of high redshift DLAs
to contain small amounts of dust similar to that found in the \caii
absorbers, the vast majority would not be included in current
magnitude-limited quasar samples.  With the reddening statistics
currently available at higher redshifts, a scenario in which 20--30\%
of DLAs have an average \EBV$\gsim0.1$, as found here for intermediate
redshifts, will not be easily distinguishable from one in which DLAs
are dust free, simply because of the substantial decrease in survey
redshift--path for the dusty absorbers.

\subsection{The number density of \caii absorbers and DLAs}\label{sec:CaDLAs}

Combining the number density per unit redshift of a class of quasar absorption
line systems, $n(z)$, with an adopted space density of systems provides a
measure of their typical cross-section on the sky.
Under the assumption of an unevolving galaxy luminosity function, the
frequency with which DLAs occur in quasar spectra implies typical
radii of $\sim$15\,$h^{-1}$\,kpc \footnote{Where $h$ is the Hubble
  constant in units of 100\,km~s$^{-1}$~Mpc$^{-1}$}; the corresponding
dimensions are $\sim$40\,$h^{-1}$\,kpc for \mgii and Lyman Limit
systems, and $\sim$70\,$h^{-1}$\,kpc for \civ absorbers
\citep[e.g.][]{1993eeg..conf..263S}.  These inferred sizes are
generally in line with the impact parameters observed directly between
galaxies and quasar sight-lines in cases where deep imaging has
revealed the galaxies responsible for the absorption systems
\citep[e.g.][]{1994AJ....108.2046S,2005ApJ...629..636A}.

Having determined $n(z)$ for \caii absorbers, we can now place them
within this broader context. Specifically, in Section \ref{sec:nz} we
deduced that the number of \caii absorbers per unit redshift with a
minimum equivalent width of 0.5\AA, $n(z,W^{{\rm lim}}=0.5)$, is
$\sim$0.013.  Correcting this value for the incompleteness due to
reddening of the background quasars as discussed above, we obtain
$n(z,W^{lim}=0.5) \gsim 0.022$ where the lower limit arises from the
unconstrained obscuration bias for systems with
\EBV$\gsim0.25$. Comparing with the values of $n(z)$ for DLAs given by
\citet{astro-ph/0505479} of $0.079 \pm 0.019$ at $0.11 < \zabs <0.9 $
and $0.120 \pm 0.025$ at $0.9 <\zabs <1.25$, we conclude that strong
\caii absorption systems have a number density of about 20--30\% that
of known DLAs.

In the simple model that different classes of absorption line systems
arise at different radii in the same galaxies, DLAs occur within the
inner regions, while \civ systems arise in outer halos, the sequence
being essentially one of decreasing neutral hydrogen column density.
The measured value of $n(z)$ for \caii absorbers then
confines them---and their associated reddening---to the innermost
7--8\,$h^{-1}$\,kpc of galaxies, where the neutral gas column density
and/or metallicity are highest.  We return to this point in
Section~\ref{sec:CaHI} below.

\subsection{Dust-to-metals ratio}\label{sec:R_DM}

We cannot deduce the gas-to-dust ratio applicable to the \caii systems
without a knowledge of their hydrogen column density.  However, having
measured metal column densities (Section~\ref{sec:metals}) and dust
content via the colour excess \EBV\ (Section~\ref{sec:redden}), we can
determine the dust-to-metals ratio typical of these absorbers.  This
ratio, which we denote ${\cal R}_{\rm DM}$, measures the degree to
which refractory elements are incorporated into dust grains which in
turn is determined by the balance between dust formation and
destruction processes.  In a recent study based on the ratio of \znii
to \feii in a sample of 38 DLAs \citet{2004A&A...421..479V} proposed
that ${\cal R}_{\rm DM}$ increases with metallicity and, more
generally, with the progress of galactic chemical evolution.  This is
in contrast to most numerical simulations which predict ${\cal R}_{\rm
DM}$ to be constant with time \citep[e.g.][]{1998ApJ...501..643D};
measurements from chemically unevolved galaxies at high redshift will
provide important constraints for such simulations in the future. At
present we simply compare the dust-to-metals ratio of the \caii
absorbers to those found in the local Universe to obtain some clue as
to their nature.

As a measure of the dust-to-metals ratio in the present study we
define ${\cal R}_{\rm DM}\equiv\langle$\EBV$ \rangle / \langle
N$(Zn~{\sc ii})$\rangle$, adopting Zn (an undepleted Fe-peak element)
as an indicator of overall metal content, and colour excess as a
substitute for dust column density. It is necessary to confine the
analysis to the subset of 27 \caii absorbers at redshifts which
include the \znii lines; their values of $N$(Zn~{\sc ii}) are given in
Table~\ref{tab:vpfit} and the lower set of values in
Table~\ref{tab:EBV} gives \EBV\ for this subset.  We make no allowance
for the incompleteness effects due to dust extinction discussed above,
since it is not possible to correct the \znii column densities without
knowing the distribution of individual values.  For the three \caii
composites, 'All', 'High-$W_{\lambda3935}$' and
'Low-$W_{\lambda3935}$', we find ${\cal R}_{\rm DM}=
8.3^{+1.9}_{-1.8}$, $6.4\pm1.2$ and $3.6^{+3.7}_{-3.6}
\times10^{-15}$\,mag\,cm$^{2}$ respectively.  The errors have been
calculated by propagation of those quoted in Tables~\ref{tab:vpfit}
and \ref{tab:EBV} and the LMC \EBV\ values have been used.  Although
at first glance there appears to be a discrepancy between the values
of the three samples, with the mean value being larger than both the
two subsamples, the errors are such that this is not significant.
Future SDSS releases, or follow up observations of \znii in individual
absorbers, may improve the situation. However, the current measurement
errors clearly do not allow us to distinguish between the 'High-' and
'Low-$W_{\lambda3935}$' samples.

We have also determined corresponding values of ${\cal R}_{\rm DM}$
appropriate to the interstellar media of the Milky Way, LMC and
SMC from their metallicities (Fe/H) and average gas-to-dust ratios
$\langle N$(H~{\sc i})$\rangle/\langle$\EBV$\rangle$, assuming a solar
(Zn/Fe) ratio \citep[$= 1.45\times10^{-3}$,][]{2003ApJ...591.1220L}
throughout:

\be
\frac {\langle \EBV \rangle}{\langle N{\rm (Zn~II)}\rangle} =
\left[ 
\frac {\langle N{\rm (H~I)}\rangle}{\langle \EBV\rangle} \times 
\frac {\rm Fe}{\rm H} \times
\left( \frac{\rm Zn}{\rm Fe}  \right)_{\odot} 
\right]^{-1}
\label{eq:EBV/Zn}
\ee
We have chosen this approach because the quantities on the right-hand
side of Eq.~\ref{eq:EBV/Zn} are much better known than the ratios 
\EBV/$N$(Zn~{\sc ii}) for individual sight-lines, especially 
to stars in the Magellanic Clouds. The results are collected in
Table~\ref{tab:dust}, where we also give references to the 
sources of the measurements. 
For the Magellanic Clouds, we have adopted the Fe abundance
determined in young B- and A-type stars---other abundance
indicators, including Zn,  are consistent with these values \citep{2004oee..symp..205H}.

\begin{table}

  \begin{center}
    \caption{\label{tab:dust} \small Gas-to-dust ratios and metal
    abundances of the Milky Way, LMC and SMC, along with derived
    dust-to-metals ratios. }
    \begin{minipage}{9.0cm}
      \begin{tabular}{cccc} \hline\hline
        & MW & LMC & SMC \\ \hline
        
        $\langle N$(H~{\sc i})$\rangle/\langle$\EBV$\rangle$ &
        0.493\footnote{\citet{1994ApJ...427..274D}} &
        2.0\footnote{\citet{1982A&A...107..247K}} &
        7.9\footnote{\citet{1990ApJS...72..163F}}\\
        
        ($10^{22}$\,cm$^{-2}$\,mag$^{-1}$) &&&\\
        
        (Fe/H) & 2.95 \footnote {\citet{2003ApJ...591.1220L}}& 
        1.12 \footnote{B-type stars, \citet{2000A&A...353..655K}} & 
        0.55 \footnote{A-type stars, \citet{1999ApJ...518..405V}} \\
        ($\times 10^{-5}$)&&&\\

        $\langle$\EBV$\rangle/\langle N$(Zn~{\sc ii})$\rangle$ & 
        4.7 & 3.1 & 1.6 \\
        
        ($10^{-15}$\,mag~cm$^{2}$) & &&\\
        
        \hline
      \end{tabular}
      \vspace*{-0.4cm}
    \end{minipage}
  \end{center}
\end{table}

The sequence of decreasing dust-to-metals ratios from the Milky Way to
the LMC and SMC evident in Table~\ref{tab:dust} is in qualitative
agreement with previous measurements \citep{1990A&A...236..237I}. It
also appears to be in accord with Vladilo's (2004) proposal of a
metallicity dependence of ${\cal R}_{\rm DM}$, although we should bear
in mind that the Fe abundance in the Magellanic Clouds has so far been
determined in only a few objects and may be subject to revision in
future.  Different star formation rates may also effect ${\cal R}_{\rm
  DM}$ \citep{1998ApJ...496..145L}.  

It is of considerable note that we find that the values determined
above for the \caii absorbers, ${\cal R}_{\rm DM} = ( 4 - 8 )
\times10^{-15}$\,mag\,cm$^{2}$, are as high, or higher, than those
derived for the Milky Way, particularly for the
'High-$W_{\lambda3935}$' sample. Previous estimates for the
dust-to-metals ratios in high redshift DLAs have been obtained by
comparison of the level of depletion of elements to that seen in the
Milky Way ISM \citep{1994ApJ...426...79P,1997ApJ...478..536P},
suggesting average ratios of $\sim$1/2 those of the Milky Way.  We saw
earlier (Fig.~\ref{fig:dla}) that the ratios of Cr, Fe and Ti to
Zn\footnote{Mn is also found to be underabundant, but this could be
due to combined dust depletion and intrinsic abundance effects
\citep{2002A&A...385..802L}.} are typically one third of solar in the
\caii absorbers, suggesting that two thirds of these refractory
elements have been incorporated into the dust.  Given that in the
Milky Way the dust depletions of these elements are $\gsim$90\%, we
would have expected values of ${\cal R}_{\rm DM}$ of about 2/3 that of
the Milky Way, i.e.$\approx 3 \times10^{-15}$\,mag\,cm$^{2}$.  Given
the uncertainties in our derivation of this ratio, the disagreement
between element depletions and dust-to-metals ratios may not be too
serious. Only further observations will clarify this issue.

\subsection{The H~{\sc i} column densities of \caii absorbers}\label{sec:CaHI}

Ultimately, we need the neutral hydrogen column densities
of the \caii absorbers in order to fully understand their relation to other
classes of quasar absorption systems. Unfortunately, this will not be
possible for at least several years, given the current lack of
space-borne UV spectrographs. We therefore have to rely
on indirect arguments based on known relations between the column
density of elements, dust content and \nhi in the local universe and
in confirmed DLAs.

Lower limits to the values of \nhi appropriate to our samples of \caii
absorbers can be derived from their \znii column densities assuming a
solar abundance of Zn. From the values in Table~\ref{tab:vpfit} we
obtain \lognhi$\ge 20.19$, 20.53 and 19.96 for the entire sample, and
'High-' and 'Low-$W_{\lambda3935}$' subsamples respectively.  Thus,
even without knowledge of their metallicity, it can be seen
immediately that the strongest ($W_{\lambda 3935}>0.68$) and, as we have shown in
Section~\ref{sec:red}, dustiest \caii systems are very likely to be
DLAs. Even the lower equivalent width sample is on average only within
a factor of two of the canonical DLA limit \lognhi\,$ = 20.3$.

The assumption of solar metallicity is likely to be overly
conservative in the present context. The abundance of Zn has been
measured in many DLAs at $z_{\rm abs} \simeq 1$ and found to be
generally sub-solar.  The recent compilation by
\citet{2005ApJ...618...68K} shows a mean and median [Zn/H]\,$\simeq
-1.0$ at these intermediate redshifts. If the \caii absorbers were
randomly chosen from known DLA samples, then their corresponding
values of \lognhi would be ten times higher than those given above,
placing them firmly at the upper end of the distribution of column
densities of DLAs.

A second constraint on \nhi can be obtained from consideration of the
gas-to-dust ratios, $\langle$ \nhi $\rangle$/$\langle$ \EBV$\rangle$,
in the Milky Way and the Magellanic Clouds given in
Table~\ref{tab:dust}.  Again, the Milky Way value will give a lower
limit to \nhi for a given reddening \EBV, if 1) the gas-to-dust ratio
increases with decreasing metallicity, as suggested by local trends
(Table~\ref{tab:dust}) and 2) the \caii absorbers have lower
metallicities than the Milky Way, as would generally be assumed for
$z\sim1$ galaxies. The values of \EBV\ in the second row of
Table~\ref{tab:EBV} and the Milky Way gas-to-dust ratio imply
\lognhi$\ge 20.38$, $20.61$, and $19.78$ for the three samples with
\znii measurements.  The corresponding values for the full sample of
37 \caii systems (top row of Table~\ref{tab:EBV}) would be somewhat
higher; correcting for the incompleteness caused by dust obscuration
up to \EBV$\lsim0.25$, as discussed in Section~\ref{sec:dustbias},
would lead to \lognhi\,$\ge 20.69$ for \caii absorbers in general.
The conclusion of this discussion is that, given their column
densities of metals and dust, \caii absorption systems with
$W_{\lambda3935} > 0.68$\,\AA\ are highly likely to be DLAs, and it is
certainly possible that all absorbers presented here are DLAs or close
to the nominal DLA column density limit.

\subsubsection{The highest column density DLAs?} \label{thehighest}
We saw earlier (Section~\ref{sec:CaDLAs}) that the number density
$n(z)$ of \caii absorbers with $W_{\lambda3935} > 0.5$\,\AA\ is
20--30\% that of DLAs after correcting for dust obscuration bias.
Possible interpretations of this statistic, together with the
conclusions above, are that \caii systems trace the subset of DLAs
with: (i) the largest values of neutral hydrogen column density; (ii)
the highest metallicities; or (iii) the largest volume densities
$n_{\rm H}$---or a combination of all three effects.  The third
possibility results from the fact that, being a minor ionisation stage
of Ca, the fraction of Ca which is singly ionised grows in proportion
to $n_{\rm H}^2$.  Thus, two clouds with the same total column density
$N$(H~{\sc i}) and the same Ca abundance, but differing volume
densities $n_{\rm H}$, will have different column densities of Ca~{\sc
ii}, due to the higher recombination coefficient for \caii in the
denser cloud \citep{1974ApJ...188L.107H}.

It is difficult to assess the relative importance of these three
possibilities without additional information, although the third one
must surely contribute, given the tendency of most \caii systems to
have stronger \mgi$\lambda 2853$ absorption for a given \mgii$\lambda
2796$ equivalent width (see Fig.~\ref{fig:mgiifeii})---neutral Mg is
also a trace state in the ISM of galaxies.

We have explored the consequences of option (i) by taking as a
starting point the column density distribution $f$[$N$(H~{\sc i})] of
DLAs in the SDSS DR3 catalogue from \citet{proch_DR3}.  With Prochaska
et al.'s double power law parameterisation for DLAs in the redshift
interval $2.5 < \zabs < 3$ the 20\% of DLAs with the largest values of
\nhi have $20.95<$\,\lognhi$<22$, and $\langle$\lognhi$\rangle=21.25$.
This value is insensitive to the exact upper limit assumed, and also
varies little with the precise redshift dependent parameterisation
used. The mean value would imply a metallicity [Zn/H]\,$= -1.06$ and a
gas-to-dust ratio $\langle$\nhi$\rangle$/$\langle$\EBV$\rangle = 3.2
\times 10^{22}$\,cm$^{-2}$~mag$^{-1}$ for the subsample of observed
\caii absorbers with spectra covering the \znii$\lambda2026$ region
and assuming LMC type dust for the latter value. In this case, the metallicity
would be typical of DLAs in general and in line with the depletion
trends in [Cr/Zn] (Fig.~\ref{fig:dla}). The gas-to-dust ratio would be
intermediate between those of the LMC and SMC (see
Table~\ref{tab:dust}).  

These parameters seem plausible, however, they are at odds with the idea
that the dust-to-metals ratio should be lower at low metallicities, as
proposed by \citet{2004A&A...421..479V}, given that in
Section~\ref{sec:R_DM} we found ${\cal R}_{\rm DM}$ to be as high, or
higher, than in the Milky Way ISM.  If the proposal of
\citet{2004A&A...421..479V} were to be correct, our detection of
reddening by the \caii absorbers argues in favour of the hypothesis
that they trace DLAs with an unusually high dust content, due to
either high metallicities or densities (or both), rather than simply
DLAs at the upper end of the distribution of $N$(H~{\sc i}). The
results of \citet{2003MNRAS.346..209L}, that the molecular hydrogen
column density of DLAs is not correlated with $N$(H~{\sc i}), but does
imply high metallicities and depletions, is also consistent with the
scenario that the dusty \caii absorbers have higher metallicities
rather than higher column densities.

\subsection{Dust in high redshift galaxies}

\begin{figure}
  \includegraphics[scale=0.6]{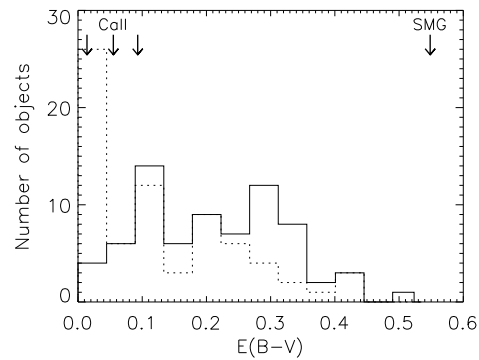}
  \caption{The distribution of \EBV\ for UV-selected galaxies at
    $z\sim2$ from \citet{2005ApJ...626..698S}. The solid and dotted
    histograms show the best-fitting \EBV\ for a constant or bursting
    star forming model respectively. Arrows indicate the average \EBV\ values
    of the observed \caii absorbers, uncorrected for incompleteness effects, 
    and the average value for the sub-millimetre selected
    galaxies of \citet{astro-ph/0507610}.}
  \label{fig:shapley}
\end{figure}

We conclude our discussion by comparing the reddening found here for
the \caii absorption-selected galaxies with measurements 
for high redshift galaxies detected directly via their stellar or dust
emission.  In two independent studies, \citet{2001ApJ...562...95S} and
\citet{2001ApJ...559..620P} concluded that Lyman break galaxies at $z
\sim 3$ have \EBV$ = 0$\,--\,$0.4$, and a median $\langle$\EBV$\rangle
\simeq 0.15$ by comparing their observed rest frame UV-to-optical
colours to model spectral energy distributions.  The addition of rest
frame near-IR photometry made possible by the \emph{Spitzer} satellite
has led to similar estimates for star-forming galaxies at $z \sim 2$
\citep{2005ApJ...626..698S} shown in Fig.~\ref{fig:shapley}. 
\citet{astro-ph/0507610} performed an equivalent analysis on a
sample of sub--millimetre selected galaxies, finding these particular
SCUBA sources to have stellar masses similar to the most massive of
the UV--selected galaxies and an average extinction in the
$V$--band, A$_V$, of 1.7 equivalent to an \EBV\ of 0.55 assuming a LMC
dust curve. This value is also indicated in the figure.

While it is not surprising to find that the highest values of \EBV\ apply to
galaxies that are luminous at far-IR/sub-mm wavelengths, where we see directly
the emission by dust, it is noteworthy that the more normal star-forming
galaxies selected via their rest frame UV stellar light are also generally more
reddened than the \caii absorbers.  An obvious question regarding the
differences in the reddening between absorption- and UV-selected galaxies
is the effect of dust obscuration bias on the former sample.  It is clearly the
case, given the calculations of dust bias presented in this chapter, that the
majority of absorbers with dust contents as high as the upper end of the
UV--selected galaxy distribution will be obscured from sight.  Whether such
absorbers represent a significant population depends primarily on the total
cross section on the sky with column densities of dust this large.  A second
complication in making a direct comparison between the effect of extinction on
the absorption- and UV-selected systems is that the reddening curves applicable
to the derivation of \EBV\ are very different for the two cases, because of the
geometrical configuration of dust and star formation in galaxies.

Nevertheless, until now, the two populations of absorption- and emission-selected
galaxies have been differentiated by dust content, as well as
metallicity and star formation rates. It is thus of potential
importance that the mean $\langle$\EBV$\rangle\gsim0.1$ which we
deduce for the \caii absorbers after correcting
for the missing systems in our magnitude limited quasar survey is not
substantially lower than the median \EBV$ = 0.15-0.20$ of star-forming
galaxies at $z = 2 - 3$. Possibly, \caii absorbers are an
intermediate link in a sequence of star formation rates, metallicities
and dust content which stretches from the relatively quiescent, metal-
and dust-poor DLAs to the actively star-forming, metal and dust-rich
Lyman break galaxies. Only further observations will provide the
information needed to confirm or refute this. 

\section{Summary and concluding remarks}

In this paper we have extended the study of intermediate redshift
($z_{\rm abs}\sim 1$) absorption systems selected via \caii equivalent
width which was begun by \citet{2005MNRAS.361L..30W} by: (a) expanding the
sample of \caii absorbers to a total of 37 with new systems selected
from the SDSS DR4; (b) measuring the relative abundances of refractory
elements from composite spectra of a subset of 27 systems at redshifts
such that the \znii$\lambda\lambda 2026, 2062$ doublet is covered in
the SDSS spectra; and (c) comparing the degree of reddening imposed on
the spectra of background quasars by dust in intervening \caii
absorbers to that caused by Mg~{\sc ii}-selected DLAs.

We have confirmed the detection of of significant reddening, at the
level of \EBV\,$\simeq 0.06$, associated with the \caii
absorption systems. This is particularly noteworthy considering that
much larger samples of Mg~{\sc ii}-selected DLAs at the same redshifts
and of Lyman~$\alpha$-selected DLAs at higher redshifts have all
failed so far to show a clear reddening signal.  Only Mg~{\sc
ii}-selected DLAs with the strongest metal lines approach the
reddening of the \caii absorbers, once allowance is made for the
fraction of sub-DLAs in the sample.

Given the lack of direct measurements of $N$(H~{\sc i}) for the
foreseeable future, we have considered various lines of indirect
evidence which all point towards the conclusion that our absorption
systems with a \caii$\lambda 3935$ equivalent width $W_{\lambda 3935}
\gsim 0.68$\,\AA\ are a subset of the DLA population, and those with
lower equivalent widths are likely to be DLAs or lie close to the
nominal DLA column density limit:  
\begin{enumerate}
\item The strengths of \mgii$\lambda 2796$ and \feii$\lambda 2600$
absorption lines fulfil the DLA selection criteria of
\citet{astro-ph/0505479} in all but two cases (which are
borderline---see Fig.~\ref{fig:mgiifeii}).
\item The ratios of the refractory elements Cr, Fe, Ti and Mn to the
undepleted Zn all span similar ranges to those in confirmed DLAs, with
the strongest \caii systems exhibiting some of the most pronounced
degrees of depletion encountered so far (Fig.~\ref{fig:dla}).
\item With the conservative assumption of solar metallicity, the
measured \znii column densities imply an average \lognhi\,$= 20.19$,
just below the threshold of \lognhi\,$= 20.3$ considered to be the
definition of a damped system. Even a moderate, and more realistic,
underabundance of Zn and other Fe-peak elements in these systems would
bring them within the conventional DLA samples.
\item In itself, the magnitude of reddening detected implies
\lognhi\,$ \ge 20.43$ adopting the conservative assumption of a
Galactic gas-to-dust ratio.
\end{enumerate}
Furthermore, these are likely to be underestimates for the population
of \caii absorbers as a whole, since our magnitude limited quasar
survey certainly misses a greater fraction of absorbers with the
largest values of \EBV.

After correcting for our detection efficiency, including dust
obscuration bias, we calculate that \caii
absorbers with $W_{\lambda 3935} > 0.5$\,\AA\ have a number density
per unit redshift $n(z) \sim 0.022$ at $\langle z_{\rm abs} \rangle =
0.95$, 20--30\% of that of Lyman~$\alpha$-selected DLAs at the same
redshifts.  This in turn implies that the regions within galaxies
giving rise to strong \caii absorption have small dimensions, with
radii of 7--8\,$h^{-1}$\,kpc. Possibly, these are the inner regions
where the highest column densities $N$(H~{\sc i}), gas densities
$n_{\rm H}$, and/or metallicities (Zn/H) are encountered and where the
characteristics of DLAs begin to approach those of the galaxies
detected directly via their UV stellar light.

Even without a knowledge of their neutral hydrogen column densities we
can deduce the typical dust-to-metals ratio $\cal R_{\rm DM}$ of
Ca~{\sc ii}-selected DLAs from their measured columns of metals (via
Zn~{\sc ii}) and dust [via \EBV].  We find high values of
$\cal R_{\rm DM}$, as high or higher than the typical ratio in the
Milky Way, although the errors affecting our estimates are
still large. This may be a further indication that \caii absorbers are
at an advanced stage of chemical evolution if the dust-to-metals ratio
correlates with metallicity.

Interestingly, we find a clear trend of increasing reddening and
element depletions with \caii equivalent width; a positive correlation
of \EBV\ with line strength is also present among the Mg~{\sc
ii}-selected DLA candidates.  Since these strong lines are generally
saturated, their equivalent widths depend more on the velocity dispersion
of the absorbers than on their column densities. Thus, these
trends may indicate that the systems with the larger values of \EBV \
are either more massive or undergoing some form of disturbance.

Whatever their exact nature, it is clear that Ca~{\sc ii}-selected
systems are an interesting class of quasar absorbers which have been
relatively neglected until now and yet merit considerable attention.
Their relation to the DLA population at large will be clarified by
measuring the \caii lines associated with confirmed DLAs; recent
improvements in detector and spectrograph efficiency at far-red and
near-IR wavelengths now make the \caii$\lambda\lambda3935,3970$ lines more easily
accessible to high resolution spectroscopy at redshifts $z_{\rm
abs} > 1$.  Such measurements will establish the relation between
$N$(H~{\sc i}), \caii equivalent width and dust content, and will
provide firm estimates of the metallicities and gas-to-dust ratios
which are lacking from the current analysis.

At $z \sim 1$ imaging is considerably easier than for the bulk of
known DLAs at $z > 2$ and it should be relatively straightforward to
identify the objects responsible for the \caii absorption. In this
respect, our sample of 37 \caii absorbers represents a major injection
of new data into a field of study which up to now has been severely
limited by the small number of known DLAs at $z \sim 1$. The images
will establish whether the absorption does take place in the inner
regions of galaxies which have undergone significant chemical
evolution, as we propose, or whether DLAs with \caii absorption single
out major mergers \citep{1991MNRAS.251..649B}. Future papers in this
series will investigate further both the relation between \caii
absorbers and DLAs, and the properties of the host galaxies of \caii
absorbers through follow-up observations.

Ultimately, the combination of absorption line spectroscopy and deep
imaging will clarify the link between absorption- and
emission-selected galaxies and thereby help constrain galaxy evolution
models. The \caii absorbers may well turn out to be the key to
understanding how different classes of high redshift galaxies fit into
a unified picture.

\section*{acknowledgments}
We would like to thank Chris Akerman, Dave Alexander, Sara Ellison and
Michael Murphy for valuable discussions, preprints of papers and
electronic data tables. VW acknowledges the award of
a PPARC research studentship. This work made extensive use of the
Craig Markwardt IDL library.

Funding for the Sloan Digital Sky Survey (SDSS) has been provided by
the Alfred P. Sloan Foundation, the Participating Institutions, the
National Aeronautics and Space Administration, the National Science
Foundation, the U.S. Department of Energy, the Japanese
Monbukagakusho, and the Max Planck Society. The SDSS Web site is
http://www.sdss.org/.  The SDSS is managed by the Astrophysical
Research Consortium (ARC) for the Participating Institutions. The
Participating Institutions are The University of Chicago, Fermilab,
the Institute for Advanced Study, the Japan Participation Group, The
Johns Hopkins University, Los Alamos National Laboratory, the
Max-Planck-Institute for Astronomy (MPIA), the Max-Planck-Institute
for Astrophysics (MPA), New Mexico State University, University of
Pittsburgh, Princeton University, the United States Naval Observatory,
and the University of Washington.

\begin{appendix}
\section{Identifying Broad Absorption Line Quasars}\label{sec:BAL}


\begin{figure*}
  \hspace*{-0.5cm}
  \begin{minipage}{16.65cm}
    \includegraphics[scale=0.75]{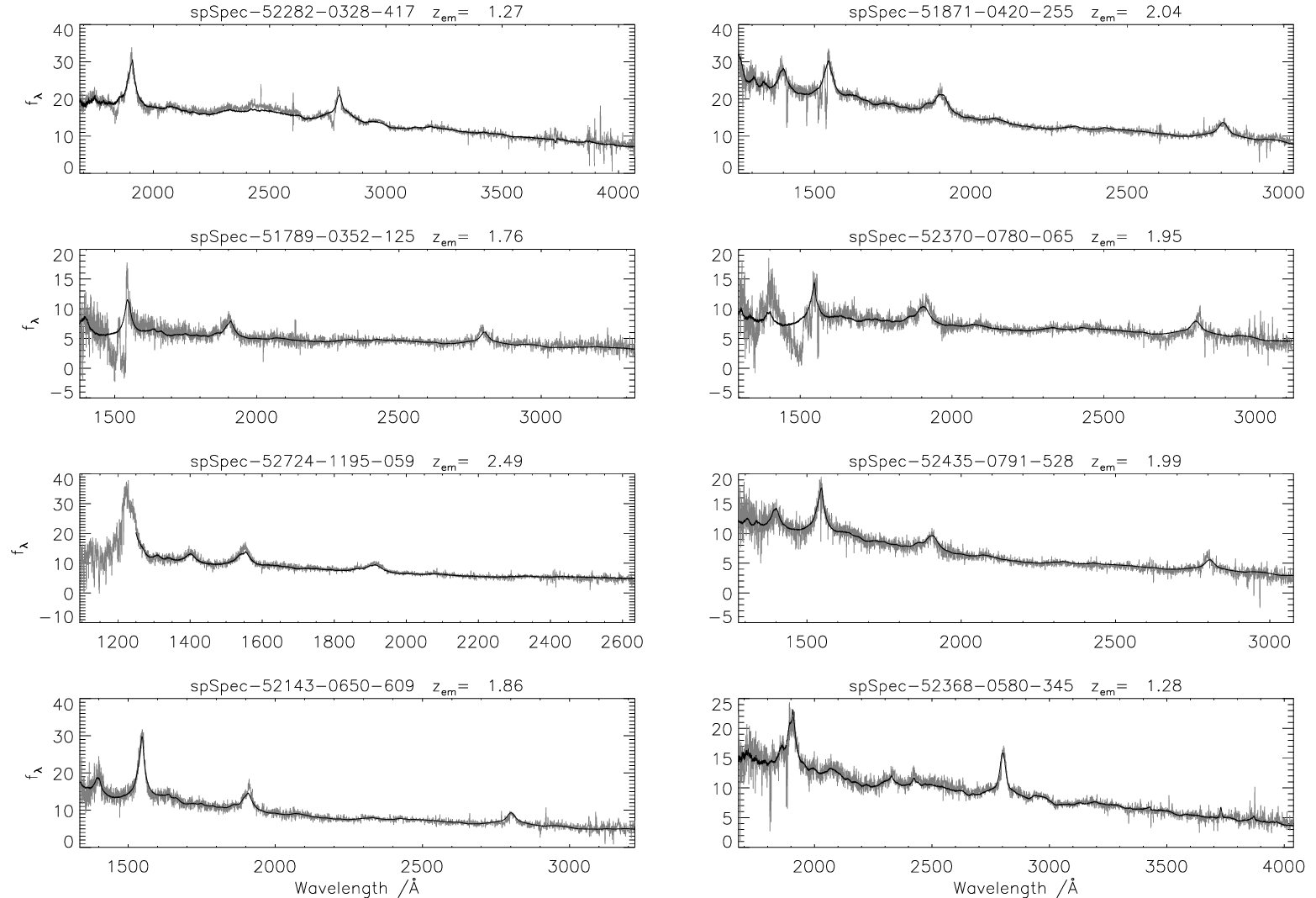}
  \end{minipage}
  \caption{Examples of quasars with (upper four panels) and without
    (lower four panels) broad absorption troughs and/or intrinsic
    absorption lines; the PCA reconstructions of the spectra are
    overplotted in black. Note the failure of the reconstructions to
    follow the majority of the BAL features due to the iterative
    removal of these objects from the sample used to create the
    eigenspectra.  Unusual quasars not used in this study are
    identified by the large $\chi^2$ between the true and
    reconstructed spectra in wavelength regions around the
    \civ$\lambda1550$, \mgii$\lambda2796$ and \feii$\lambda2600$
    lines.  The $y$-axis is in units of
    $10^{-17}$\,erg~cm$^{-2}$~s$^{-1}$~\AA$^{-1}$.  The SDSS
    spectroscopic filename (Modified Julian Date, plate number, fibre
    number) and quasar redshift are given above each panel.}
  \label{fig:bals}
\end{figure*}

BAL quasars are known to make up around 15\% of all active galactic
nuclei at $\zem > 1$ \citep{2003AJ....125.1784H,2003AJ....126.2594R}.
They are characterised by broad absorption troughs blueward of
emission lines of many ionisation states---presumably arising in
outflowing material---and an overall reddening of the quasar spectral
energy distribution.  While much effort is being devoted to
understanding the origin of these spectral features and the position
of BAL quasars in the unified model of active galactic nuclei, their
generally redder SEDs are a potential source of additional uncertainty
in studies of reddening caused by intervening absorbers. Similarly
associated absorption systems, in which hydrogen and metal absorption
lines seen in the spectra of quasars are associated with the quasar
host galaxy itself, also appear in quasars with generally redder SEDs
\citep{1998ApJ...494..175C}. Thus, we need to remove quasars whose
spectra show evidence of BALs, or strong associated absorption, from
the sample prior to our dust reddening analysis.  Due to the large number of quasar
spectra in the SDSS, the problem becomes one of developing suitable
automatic pattern recognition techniques capable of coping with the
wide range in strength and shape of the absorption features.

The main difficulty lies in defining the quasar continuum around the emission
lines.  Traditionally, this has been achieved by fitting power laws to the
spectra away from regions containing emission lines.
\citet{2003AJ....125.1711R} compared this method to the use of composite spectra
in the SDSS early data release; they also provide a history of BAL detection
and an extensive discussion of the different methods which will not be repeated
here.  Here, we present a third method which makes use of a principal
component analysis (PCA) of the spectra to reconstruct the quasar continua and
emission lines, without the broad absorption features.  The main advantage of
PCA over the use of composite spectra for reconstructing the quasar continua is
the tailoring of the fit to each individual object; all the detailed
information which is lost through power law fits can also be retained.

\subsection{Principal Component Analysis}

PCA has become a standard spectral analysis technique in astronomy,
generally used to automatically classify stars, galaxies and quasars
\citep[e.g.][]{1992ApJ...398..476F, 1995AJ....110.1071C,
2002MNRAS.333..133M, 2004AJ....128..585Y}, but also in data reduction
\citep{1998ApJ...492...98G,skysub} and reconstruction problems.
Mathematical details can be found in standard texts on multivariate
analysis \citep[e.g.][]{1975kendall,1987mda..book.....M}, but the main
aspects of the technique can be easily explained.  An array of $N$
spectra each containing $M$ pixels can alternatively be visualised as
each spectrum being a point in an $M$ dimensional space.  PCA searches
for the lines of greatest variance in the cloud of points representing
the spectra, each line being orthogonal to all the previous ones,
therefore picking out a basis set in which the eigenvectors (principal
components or `eigenspectra') are ordered according to their relative
contribution to the overall variance in the array.  PCA has no
knowledge of physical information; consequently physical features
which occur in only a few spectra will be lost in the noise and
`features' resulting from low SNRs may be erroneously identified as
signal.  The former property is advantageous when identifying the
small fraction of spectra exhibiting strong absorption.

Once the eigenspectra are created, any suitable quasar spectrum can be
reconstructed by projection onto the eigenspectra.  The reconstructed spectrum
will only contain the components present in the eigenspectra and a poor fit will
be obtained for quasar spectra with shape and/or features not well represented
in the input sample.

\subsection{Details of the method}\label{subsec:BAL_method}

Given the size of the SDSS catalogue, it is possible to create high
SNR eigenspectra for different subsamples of quasars.  Splitting the
sample into redshift bins can greatly simplify the PCA analysis by
minimising the `missing data' caused by differing rest wavelength
coverage.  We defined three redshift bins such that all the spectra
contained within them have less than 25\% of their pixels missing when
placed onto a common rest frame wavelength array: $0.823<\zem<1.537$,
$1.405<\zem<2.179$ and $2.172<\zem<3.193$.  We chose to have a slight
overlap between the redshift bins in order to increase the number of
spectra contributing to either end of the eigenspectra, and thus
improve the SNR at the ends of the eigenspectra; a small fraction of the spectra
therefore contribute to two bins.  We also limited our analysis to
wavelengths longwards of 1250\,\AA, i.e. away from the Lyman-$\alpha$
forest and the peak of the Lyman-$\alpha$ emission in the quasars.

Several steps are involved before the final sample of BAL quasars was
produced.  The first step was to identify strong BAL quasar candidates
using a simple ``colour cut'', based on the difference in the flux
contained within bandpasses on either side of the \civ$\lambda1550$
emission line (or the \mgii$\lambda2796$ line when \civ is not
covered); these obvious BAL quasars were then temporarily removed from
the sample.  The remaining spectra were shifted to the rest frame
(without rebinning) and a PCA performed\footnote{Because of the large
number of pixels in the SDSS spectra, we employed an expectation
maximisation algorithm \citep{EMpca} which affords a substantial
increase in speed.}  on the spectra in each redshift bin using the
``gappy-PCA'' method of \citet{1999AJ....117.2052C}.  Gappy-PCA is an
iterative procedure in which missing pixels in each
spectrum\footnote{Pixels can be ``missing'' from a spectrum either
because of defects or intrinsic rest frame wavelength coverage.} are
replaced by the reconstructed spectra created from the eigenspectra of
the previous iteration.  During reconstruction each pixel is weighted
by the inverse of the squared error array, with missing pixels given
zero weight.  

A 41-pixel median filter was run through each quasar spectrum to
identify narrow absorption features. The spectra were then
reconstructed from the eigenspectra with these narrow features masked
to prevent them from biasing the reconstructed continua
low\footnote{Masks were applied during the gappy-PCA procedure by
giving the pixels zero weight.}.  Reconstructed continua were created
for the entire sample of quasars, including those excluded during the
creation of the eigenspectra by the colour cuts around the
\civ$\lambda1550$ and \mgii$\lambda2796$ lines.  For those interested
in the ``BALnicity'' of individual objects
\citep[e.g.][]{1991ApJ...373...23W}, the quasar spectra are best
reconstructed using only those eigenspectra visually identified as
containing no BAL features, with the regions in which BAL features are
expected masked during reconstruction so as not to bias the continua
too low.  Fig.~\ref{fig:bals} shows examples of reconstructions for
both BAL and non-BAL quasar spectra.

The number of eigenspectra used in a reconstruction is not defined by
the PCA procedure. For our purposes of identifying and removing
unusual spectra, the precise number was not critical: BAL spectra and
quasar spectra with associated absorption possess large $\chi^2$
values between the true and reconstructed spectrum in the region of
BAL features even when as many as 10-15 eigenspectra are used for
reconstruction. This reflects the large variations in size and shape
of BAL features which the PCA is unable to reconstruct even with
relatively large numbers of components.  As we were not interested in
the precise classification of each object, we simply plotted the
distribution of $\chi^2$ between the true and reconstructed spectra in
wavelength regions blueward of the \civ$\lambda1550$,
\mgii$\lambda2796$ and/or \feii$\lambda2600$ lines.  All objects which
fell into the high $\chi^2$ tail of each distribution were removed,
producing a new input sample for a second iteration of the PCA.  The
entire PCA and reconstruction procedure was repeated twice to ensure
that as few BALs as possible were present when creating the final
eigenspectra and the eigenspectra were therefore most representative
of ordinary quasars.  In total, 13\% of quasars, with redshifts $0.85
\le z \le 3.2$, were removed from the input sample.

Ultimately, our aim was to create a subsample of quasars suitable for:
(a) searching for intervening absorption systems and (b) creating
control spectra with which to compare the SEDs of quasars with
absorbers.  The final results reported in this paper are insensitive
to the details of the scheme adopted to identify and exclude BAL
quasars.  By inadvertently removing any non-BAL quasars from the
sample we would be reducing the sightlines available for the search,
but this is a small effect given the size of the input sample.  On the
other hand, any BAL quasars which remain in the sample will increase
the variance among the quasar SEDs, so reducing our ability to isolate
any systematic differences between the SEDs of quasars with
intervening metal absorption line systems and the quasar population as
a whole.  This effect is included in the Monte Carlo simulations used
to estimate the mean and variance of SED colours for random
samples of quasars (see Section \ref{subsec:mc}).

\end{appendix}

\end{document}